\documentclass[11pt,letterpaper]{article}

\usepackage[letterpaper,margin=1in]{geometry}
\usepackage[T1]{fontenc}
\usepackage{times}
\usepackage{microtype}
\usepackage[hyphens]{url}
\usepackage{graphicx}
\usepackage{amsmath,amssymb,amsfonts}
\usepackage{booktabs}
\usepackage{multirow}
\usepackage{array}
\usepackage{xcolor}
\usepackage{natbib}
\usepackage{caption}
\usepackage{float}
\usepackage{algorithm}
\usepackage{algorithmic}
\usepackage{placeins}
\usepackage{hyperref}

\hypersetup{
    colorlinks=true,
    linkcolor=blue,
    citecolor=blue,
    urlcolor=blue,
    pdftitle={DY-LUT: Depth-Aware YCbCr Lookup Tables for Real-Time Underwater Image Enhancement},
    pdfauthor={Cunhao Zhu, Xiangtao Kong, Dongliang Xu, Zhiheng Zhang, Tianyu Wang, Yue Yao}
}
\newcommand{\best}[1]{\textbf{#1}}
\newcommand{\second}[1]{\underline{#1}}
\newcommand{\third}[1]{\textit{#1}}
\newcommand{\CDIone}{\mathrm{CDI}_1}
\newcommand{\CDItwo}{\mathrm{CDI}_2}
\providecommand{\Description}[1]{}

\title{DY-LUT: Depth-Aware YCbCr Lookup Tables for Real-Time Underwater Image Enhancement}
\author{%
Cunhao Zhu\textsuperscript{1} \quad
Xiangtao Kong\textsuperscript{2} \quad
Dongliang Xu\textsuperscript{1}\\
Zhiheng Zhang\textsuperscript{1} \quad
Tianyu Wang\textsuperscript{3} \quad
Yue Yao\textsuperscript{1}\\[0.6em]
\small \textsuperscript{1}Shandong University\\
\small \textsuperscript{2}The Hong Kong Polytechnic University\\
\small \textsuperscript{3}Mohamed bin Zayed University of Artificial Intelligence
}
\date{}

\begin{document}
\maketitle

\begin{abstract}
Underwater image enhancement is challenged by spatially non-uniform, wavelength-dependent attenuation. Propagation distance and wavelength govern this degradation, while YCbCr separates luminance from chrominance for restoration. We propose DY-LUT, a depth-aware YCbCr lookup-table framework for real-time enhancement. A dual-branch encoder predicts image-level fusion weights and a joint pair of pixel-wise degradation indices from image and depth features. These quantities condition learnable 4D LUTs, followed by lightweight local refinement. DY-LUT preserves traditional LUT efficiency while enabling depth-conditioned, spatially adaptive restoration. With externally supplied depth, its 3.56M-parameter enhancement network achieves competitive quality on UIEB-90 and LSUI and runs $9$--$304\times$ faster than representative high-capacity baselines. Adaptive inference further maintains real-time performance ($\sim7$ ms) for 4K UIQAD images. DY-LUT also benefits downstream detection and feature matching. Ablations show that YCbCr is a more effective basis than RGB for depth-conditioned lookup, while the jointly learned indices further improve adaptive querying. These results provide a physically grounded route to efficient UIE on practical platforms.
\end{abstract}

\section{Introduction}

Underwater vision supports ocean exploration, autonomous navigation, and ecological monitoring, where image enhancement directly affects subsequent perception. Practical systems need both restoration quality and low-latency high-resolution processing on constrained platforms.

\begin{figure}[t]
    \centering
    \includegraphics[width=\linewidth]{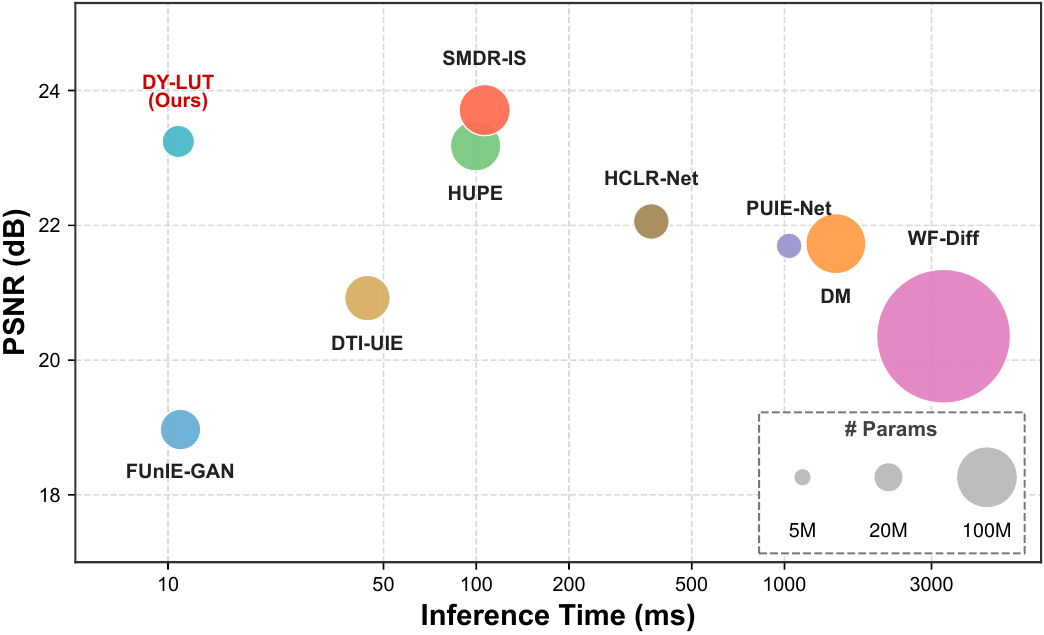}
    \caption{UIEB-90 enhancement quality--efficiency and model size with externally supplied depth; RGB-only end-to-end cost is reported in Sec.~\ref{sec:DepthAnalysis}.}
    \Description{A bubble chart comparing underwater image enhancement methods on the UIEB-90 benchmark in terms of PSNR, inference time, and model size. Bubble area indicates the number of parameters.}
    \label{fig:fig1}
\end{figure}

Underwater images suffer from wavelength-dependent absorption and scattering, insufficient illumination, and sensor noise. Unequal spectral attenuation causes color shifts, contrast degradation, and detail loss, while scene depth---the object-to-sensor distance considered in this work---makes these degradations spatially non-uniform \cite{saoud2024seeing,akkaynak2018revised,schechner2006recovery,zhou2023multicolor}. Effective UIE must therefore model wavelength and scene depth jointly.

\newpage
Scene depth cues propagation distance and attenuation strength. YCbCr complements it by separating luminance and chrominance, facilitating targeted modeling of brightness loss and color distortion. Together, they organize wavelength-dependent, spatially varying degradation.

Based on this insight, we propose \textbf{DY-LUT}, a depth-aware YCbCr lookup table for real-time UIE. A dual-branch encoder predicts image-level LUT fusion weights and a joint two-dimensional field of degradation indices from the degraded image and its depth map. These predictions condition a bank of learnable 4D LUTs in YCbCr space. Local refinement then improves spatial consistency.

Our contributions are threefold. First, we formulate depth-conditioned lookup in YCbCr space, using depth to represent propagation-distance variation and YCbCr to organize luminance and chrominance restoration. Second, we introduce a dual-branch encoder, a learnable 4D LUT bank, and local refinement in a compact 3.56M-parameter framework; the two degradation indices are optimized jointly as latent lookup coordinates without assigning separate physical semantics. Third, we evaluate full-reference quality, zero-shot generalization, high-resolution efficiency, depth-source robustness, and downstream perception under a unified protocol.

Experiments on UIEB-90 and LSUI show a favorable quality--efficiency balance (Fig.~\ref{fig:fig1}), including a $9$--$304\times$ speedup over representative high-capacity baselines. On an NVIDIA RTX 4090D, adaptive inference takes about $7$\,ms for 4K UIQAD images. U45, UIEB-C60, detection, and matching results further support practical generalization.

\section{Related Work}

\noindent\textbf{Deep Learning Based Underwater Enhancement.}
Early deep UIE methods use adversarial learning to correct color and contrast~\cite{fabbri2018enhancing,islam2020fast}, while later approaches incorporate uncertainty, multiple color spaces, semantic guidance, and underwater priors~\cite{fu2022uncertainty,wang2021uiec,qi2022sguie,li2020underwater}. Recent Transformer, state-space, frequency-domain, diffusion, contrastive, and downstream-task-aware models further improve restoration quality~\cite{peng2023u,guan2024watermamba,zhao2024wavelet,zhang2024synergistic,zhou2024hclr,zhang2025hupe,lin2026dtiuie}, but their cost often precludes real-time use.

\noindent\textbf{LUT-based Image Enhancement.}
LUT methods combine offline learning with efficient online lookup. Image-adaptive 3D LUTs~\cite{zeng2020learning} have been extended with spatial awareness, adaptive sampling, compressed or implicit parameterization, and content-aware 4D lookup~\cite{wang2021real,yang2022adaint,yang2022seplut,zhang2022clut,conde2024nilut,liu20234d}. INAM~\cite{xiao2023inam} demonstrates the efficiency of LUTs for UIE, but existing underwater designs largely remain RGB-based 3D mappings with limited capacity for depth-dependent degradation.

\noindent\textbf{Depth-aware Underwater Enhancement.}
Scene depth can be obtained from onboard sensors, stereo systems, or monocular estimators~\cite{yang2024depth,yang2024depthv2,zhang2024atlantis}. Sea-Thru~\cite{akkaynak2019sea} and IBLA~\cite{peng2017underwater} use depth-related cues for physical inversion, whereas learning-based frameworks employ depth or transmission as an input, auxiliary modality, or optimization target~\cite{li2021underwater,bar2024osmosis,huang2025underwater,du2024physical}. These studies motivate combining explicit depth cues with a color representation tailored to wavelength-dependent underwater degradation.

\begin{figure*}[t]
    \centering
    \includegraphics[width=\textwidth]{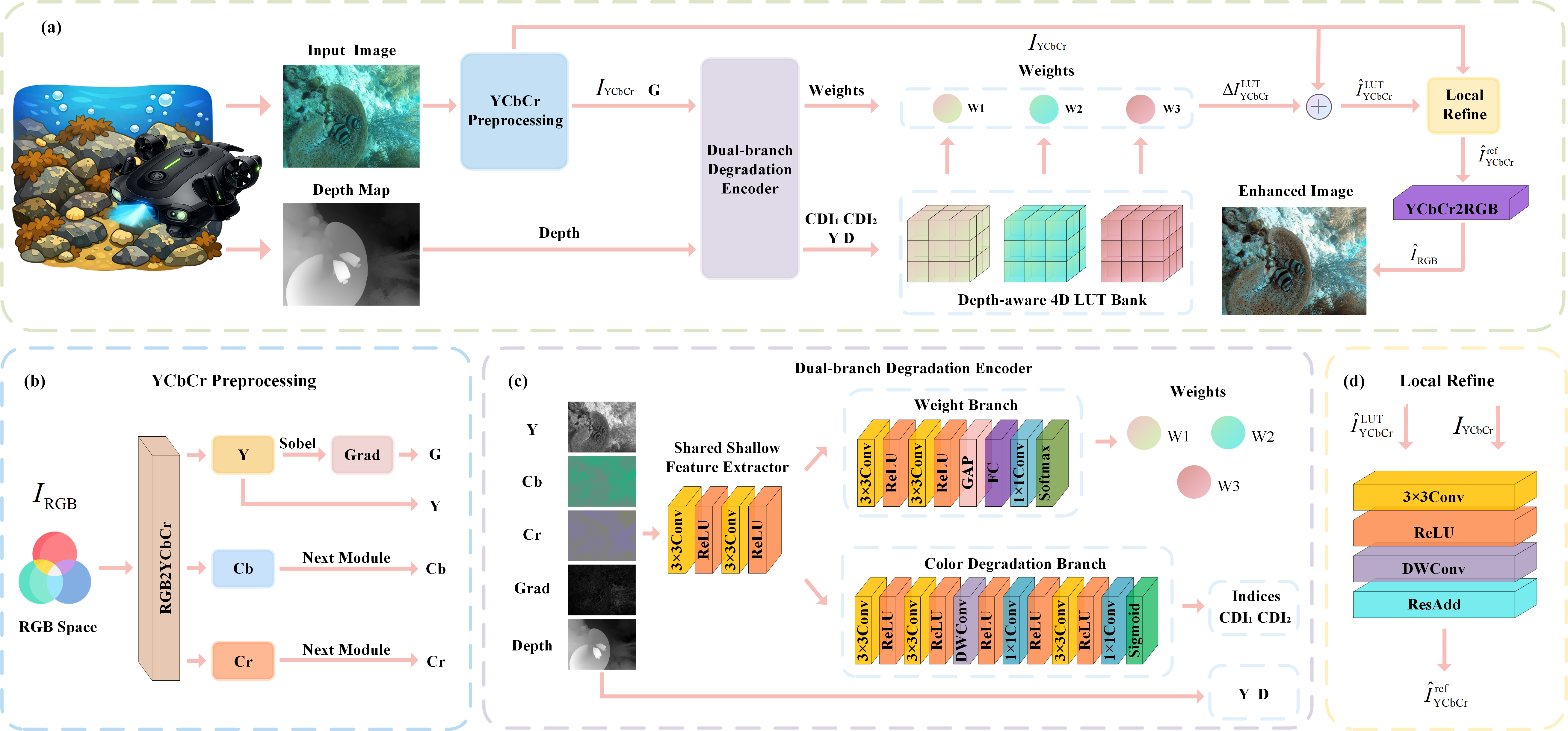}
    \caption{DY-LUT overview: (a) pipeline, (b) YCbCr preprocessing, (c) dual-branch encoder, and (d) local refinement.}
    \Description{Overall framework of DY-LUT, including YCbCr Preprocessing, degradation encoder, a depth-aware LUT bank, and a local refinement module.}
    \label{fig:framework}
\end{figure*}

\section{Methodology}

\subsection{Motivation}

Underwater degradation depends strongly on scene depth and wavelength. According to the Beer--Lambert law~\cite{akkaynak2018revised}, the direct-transmission component decays exponentially with propagation distance:
\begin{equation}
E_{\mathrm{dir}}(\lambda,D)
=E_0(\lambda)t(\lambda,D),\qquad
t(\lambda,D)=e^{-\mu(\lambda)D},
\end{equation}
where \(E_{\mathrm{dir}}\) and \(E_0\) denote the attenuated direct component and its unattenuated value, \(\lambda\) is wavelength, \(D\) is scene depth, and \(\mu(\lambda)\) is the wavelength-dependent attenuation coefficient. The complete observation may also contain backscatter; the equation isolates direct attenuation to motivate our representation rather than serving as a full image-formation model. Because red light decays faster than blue-green light, increasing \(D\) produces both luminance attenuation and wavelength-dependent color shifts. In RGB, reflectance and unequal channel attenuation remain coupled, making brightness degradation and color distortion difficult to organize separately.

We therefore transform the input image \(I_{\mathrm{RGB}}\) into \(I_{\mathrm{YCbCr}}=[Y,C_b,C_r]\) using the normalized ITU-R BT.601 conversion~\cite{itu2011bt601}. The luminance channel \(Y\) characterizes energy attenuation and structural visibility, while \(C_b\) and \(C_r\) represent chromatic variation. Combining this separation with the depth cue \(D\) yields a compact representation of depth-dependent, wavelength-selective degradation, thereby motivating the DY-LUT design (Tab.~\ref{tab:ablation_rep}).

\subsection{Overview}
\label{sec:Overview}

As illustrated in Fig.~\ref{fig:framework}, DY-LUT first converts the input \(I_{\mathrm{RGB}}\) to \(I_{\mathrm{YCbCr}}=[Y,C_b,C_r]\) and extracts the luminance gradient \(\Gamma=\mathrm{Sobel}(Y)\)~\cite{sobel1968a}. The encoder input \(\mathbf{X}=\mathrm{cat}[Y,C_b,C_r,D,\Gamma]\) is processed by a shared extractor and two branches that predict image-level LUT fusion weights \(\boldsymbol{\alpha}\) and a joint pixel-wise index field \(\mathbf{C}_{\mathrm{DI}}=[\CDIone,\CDItwo]\). A 4D LUT bank uses \((Y,D,\CDIone,\CDItwo)\) to predict a YCbCr residual, and a lightweight refinement module corrects local inconsistencies before conversion back to the final RGB output \(\hat I_{\mathrm{RGB}}\). The modules are detailed below.

\subsection{Dual-branch Degradation Encoder}
\label{sec:DualBranchEncoder}

The Dual-branch Degradation Encoder in Fig.~\ref{fig:framework}(c) extracts the shared feature \(\mathbf{F}_{s}=f_{\mathrm{s}}(\mathbf{X})\), which is passed to a Weight Branch for image-level LUT fusion and a Color Degradation Branch for pixel-wise index prediction.

The encoder receives five channels, \(Y,C_b,C_r,D,\Gamma\), but keeps the lookup 4D for a favorable capacity--efficiency trade-off. A 5D LUT would increase its parameters by more than twenty times~\cite{zeng2020learning,liu20234d,xiao2023inam}. The gradient \(\Gamma\) supports feature extraction and index prediction rather than defining another lookup axis.

\subsubsection{Weight Branch}

The Weight Branch aggregates the shared feature into a global representation \(\mathbf{z}\) and predicts the image-level LUT fusion weights \(\boldsymbol{\alpha}\):
\begin{equation}
\mathbf{z}=\mathrm{Pool}(\mathbf{F}_{s}), \qquad
\boldsymbol{\alpha}=\mathrm{Softmax}\!\left(\mathrm{Head}_{\mathrm{w}}(\mathbf{z})\right),
\end{equation}
where \(\boldsymbol{\alpha}=[\alpha_1,\dots,\alpha_K]\), \(\sum_{k=1}^{K}\alpha_k=1\), and \(K=3\) following prior adaptive LUT designs~\cite{zeng2020learning,liu20234d}.

\subsubsection{Color Degradation Branch}

This branch predicts a joint two-channel index field, \(\mathbf{C}_{\mathrm{DI}}=[\CDIone,\CDItwo]\), whose channels supply the third and fourth lookup dimensions:
\begin{equation}
\mathbf{C}_{\mathrm{DI}}=\sigma\!\left(\mathrm{Head}_{\mathrm{cd}}(\mathbf{F}_{s})\right)
\in[0,1]^{N_{\mathrm b}\times 2\times H\times W},
\end{equation}
where \(\sigma(\cdot)\) is the Sigmoid activation, \(N_{\mathrm b}\) is the batch size, and \(H,W\) are the image height and width. The two channels are optimized jointly as latent lookup coordinates. Because no loss assigns them distinct physical meanings, we do not interpret or ablate either channel in isolation; Tab.~\ref{tab:ablation_rep} evaluates the pair against fixed chrominance coordinates.

\subsection{Depth-aware 4D LUT Bank}
\label{sec:DepthAware4DLUT}

The bank contains \(K=3\) learnable LUTs. Each \(\mathcal{T}_k\in\mathbb{R}^{N\times N\times N\times N\times3}\) stores a YCbCr residual in its final dimension, and every lookup axis uses \(N=25\). After normalizing and clamping the four coordinates to \([0,1]\), quadrilinear interpolation combines the 16 neighboring vertices using separable linear weights along \((Y,D,\CDIone,\CDItwo)\). The exact interpolation equation and pseudocode are provided in the supplement. The weighted residual and preliminary result are defined as follows:
\begin{equation}
\begin{aligned}
\Delta I_{\mathrm{YCbCr}}^{\mathrm{LUT}}
&=\sum_{k=1}^{K}\alpha_k\,
\mathrm{Interp}\!\left(\mathcal{T}_k;Y,D,\CDIone,\CDItwo\right),\\
\hat I_{\mathrm{YCbCr}}^{\mathrm{LUT}}
&=I_{\mathrm{YCbCr}}+\Delta I_{\mathrm{YCbCr}}^{\mathrm{LUT}}.
\end{aligned}
\end{equation}

\subsection{Local Refinement Module}

The proposed Local Refinement Module is illustrated in Fig.~\ref{fig:framework}(d). Since LUT-based residual is spatially independent (pixel by pixel), it may introduce local inconsistencies near edges or in highly textured regions. To alleviate this issue, we further introduce a lightweight refinement module in the same YCbCr space.

The refinement input concatenates the preliminary result and the original YCbCr image. It then predicts a correction:
\begin{equation}
\begin{aligned}
\mathbf{X}_{\mathrm{ref}}
&=\mathrm{cat}[\hat I_{\mathrm{YCbCr}}^{\mathrm{LUT}},I_{\mathrm{YCbCr}}],\\
\Delta I_{\mathrm{YCbCr}}^{\mathrm{ref}}
&=f_{\mathrm{ref}}(\mathbf{X}_{\mathrm{ref}}),\\
\hat I_{\mathrm{YCbCr}}^{\mathrm{ref}}
&=\hat I_{\mathrm{YCbCr}}^{\mathrm{LUT}}+\Delta I_{\mathrm{YCbCr}}^{\mathrm{ref}}.
\end{aligned}
\end{equation}
Finally, \(\hat I_{\mathrm{YCbCr}}^{\mathrm{ref}}\) is converted back to the RGB output \(\hat I_{\mathrm{RGB}}\).

\begin{figure*}[!t]
    \centering
    \includegraphics[width=\textwidth]{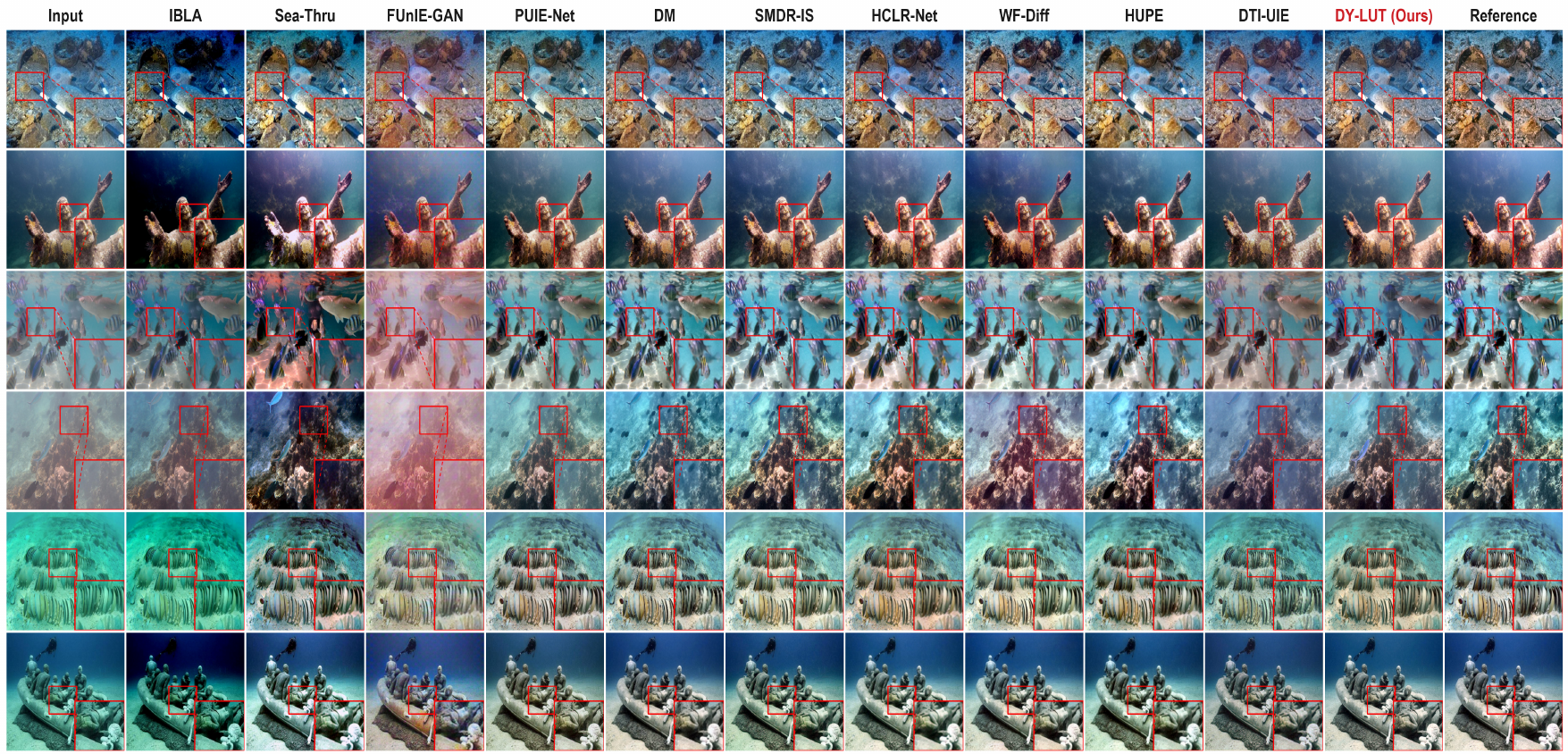}
    
    \caption{Qualitative comparisons on near-, far-, mixed-depth, and cross-dataset scenes.}
    \Description{Qualitative comparison of underwater image enhancement methods on representative scenes, including input images, ground truth when available, several competing methods, and DY-LUT.}
    \label{fig:qualitative}
\end{figure*}

\begin{table*}[!t]
\centering
\scriptsize
\setlength{\tabcolsep}{0.22mm}
\renewcommand{\arraystretch}{1.00}
\resizebox{0.97\textwidth}{!}{%
\begin{tabular}{@{}cllccccccccccc@{}}
\toprule
\multirow{2}{*}{\textbf{Type}} & \multirow{2}{*}{\textbf{Method}} & \multirow{2}{*}{\textbf{Venue}} & \multirow{2}{*}{\textbf{Space}} & \multirow{2}{*}{\textbf{Params (M)}$\downarrow$}
& \multicolumn{5}{c}{\textbf{UIEB-90}}
& \multicolumn{4}{c}{\textbf{LSUI}} \\
\cmidrule(lr){6-10} \cmidrule(l){11-14}
& & & &
& \textbf{PSNR}$\uparrow$ & \textbf{SSIM}$\uparrow$ & \textbf{LPIPS}$\downarrow$ & \textbf{Time (ms)}$\downarrow$ & \textbf{FLOPs (G)}$\downarrow$
& \textbf{PSNR}$\uparrow$ & \textbf{SSIM}$\uparrow$ & \textbf{LPIPS}$\downarrow$ & \textbf{Time (ms)}$\downarrow$ \\
\midrule
\multirow{2}{*}{Physics}
& IBLA & TIP'17 & RGB & ---
& 14.48 & 0.6153 & 0.3555 & \third{11.55} & \best{1.06}
& 15.18 & 0.6097 & 0.3895 & \second{5.92} \\
& Sea-Thru & CVPR'19 & RGB & ---
& 11.26 & 0.5235 & 0.5311 & 87.63 & \second{6.38}
& 11.42 & 0.5081 & 0.5393 & 45.23 \\
\midrule
\multirow{9}{*}{DL}
& FUnIE-GAN & RA-L'20 & RGB & 7.02
& 18.969 & 0.7834 & 0.2302 & \second{10.97} & 68.4
& 18.058 & 0.7583 & 0.2917 & \best{3.11} \\
& PUIE-Net & ECCV'22 & RGB & \best{1.40}
& 21.695 & 0.8941 & \third{0.1259} & 1035 & 8362
& \third{20.995} & \third{0.8249} & \third{0.2306} & 194.3 \\
& DM & MM'23 & RGB & 18.3
& 21.728 & \best{0.9047} & \second{0.1254} & 1468 & 14833
& 18.157 & 0.7757 & 0.2765 & 132.4 \\
& SMDR-IS & AAAI'24 & RGB & 12.6
& \best{23.710} & \third{0.8959} & \best{0.1203} & 106.5 & 913.3
& 20.609 & 0.8216 & 0.2323 & 40.19 \\
& HCLR-Net & IJCV'24 & RGB & \third{4.87}
& 22.057 & 0.8663 & 0.1415 & 369.9 & 7116
& 20.748 & \second{0.8263} & \best{0.2193} & 60.97 \\
& WF-Diff & CVPR'24 & RGB & 100.6
& 20.354 & 0.8787 & 0.1595 & 3283 & 26339
& 19.779 & 0.7679 & 0.2633 & 296.0 \\
& HUPE & IJCV'25 & RGB & 12.18
& \third{23.18} & 0.8930 & 0.1344 & 99.50 & 153.60
& 20.79 & 0.8189 & 0.2680 & 92.93 \\
& DTI-UIE & TIP'26 & RGB & 9.5
& 20.920 & 0.8608 & 0.1501 & 44.35 & 361
& \second{21.09} & 0.8175 & 0.2364 & 43.52 \\
& \textbf{DY-LUT} & None & YCbCr & \second{3.56}
& \second{23.245} & \second{0.8994} & 0.1351 & \best{10.79} & \third{21.11}
& \best{21.14} & \best{0.8271} & \second{0.2217} & \third{6.52} \\
\bottomrule
\end{tabular}%
}
\caption{UIEB-90 and zero-shot LSUI comparison. Quality metrics use \(256\times256\); enhancement efficiency uses native resolution and excludes depth acquisition. FLOPs appear once under UIEB-90; styles mark the top three results.}
\label{tab:main_results}
\end{table*}

\subsection{Training Objective}

We train DY-LUT end-to-end in its operating color space. Let \(I_{\mathrm{gt}}\) denote the RGB ground truth. Converting \(\hat I_{\mathrm{RGB}}\) and \(I_{\mathrm{gt}}\) to YCbCr gives \((\hat Y,\hat C_b,\hat C_r)\) and \((Y_{\mathrm{gt}},C_{b,\mathrm{gt}},C_{r,\mathrm{gt}})\), respectively. The objective is
\begin{equation}
\begin{aligned}
\mathcal{L}_{\mathrm{rec}}={}&
\|\hat Y-Y_{\mathrm{gt}}\|_1
+\beta_{C_b}\|\hat C_b-C_{b,\mathrm{gt}}\|_1\\
&+\beta_{C_r}\|\hat C_r-C_{r,\mathrm{gt}}\|_1\\
&+\beta_{\mathrm{SSIM}}
\bigl(1-\mathrm{SSIM}(\hat Y,Y_{\mathrm{gt}})\bigr),\\
\mathcal{L}_{\mathrm{total}}={}&\mathcal{L}_{\mathrm{rec}}
+\beta_{\mathrm{TV}}\mathcal{L}_{\mathrm{TV}}
+\beta_{\mathrm{MN}}\mathcal{L}_{\mathrm{MN}}\\
&+\beta_{\mathrm{VGG}}\mathcal{L}_{\mathrm{VGG}}
+\beta_{\mathrm{grad}}\mathcal{L}_{\mathrm{grad}}.
\end{aligned}
\end{equation}
Here, \(\mathcal{L}_{\mathrm{TV}}\) smooths neighboring LUT entries and \(\mathcal{L}_{\mathrm{MN}}\) preserves order along the luminance lookup dimension~\cite{zeng2020learning,liu20234d}. The perceptual term \(\mathcal{L}_{\mathrm{VGG}}\) is computed in RGB using pretrained VGG features~\cite{simonyan2015very,johnson2016perceptual}, while \(\mathcal{L}_{\mathrm{grad}}\) preserves luminance edges. We set \(\beta_{C_b}=\beta_{C_r}=1.5\), \(\beta_{\mathrm{SSIM}}=1.0\), \(\beta_{\mathrm{TV}}=5\times10^{-5}\), \(\beta_{\mathrm{MN}}=2.0\), \(\beta_{\mathrm{VGG}}=0.1\), and \(\beta_{\mathrm{grad}}=0.05\).

The architecture and loss ablations are reported together, with their results summarized in Tab.~\ref{tab:ablation_arch_loss}.

\begin{figure*}[!t]
    \centering
    \includegraphics[width=\textwidth]{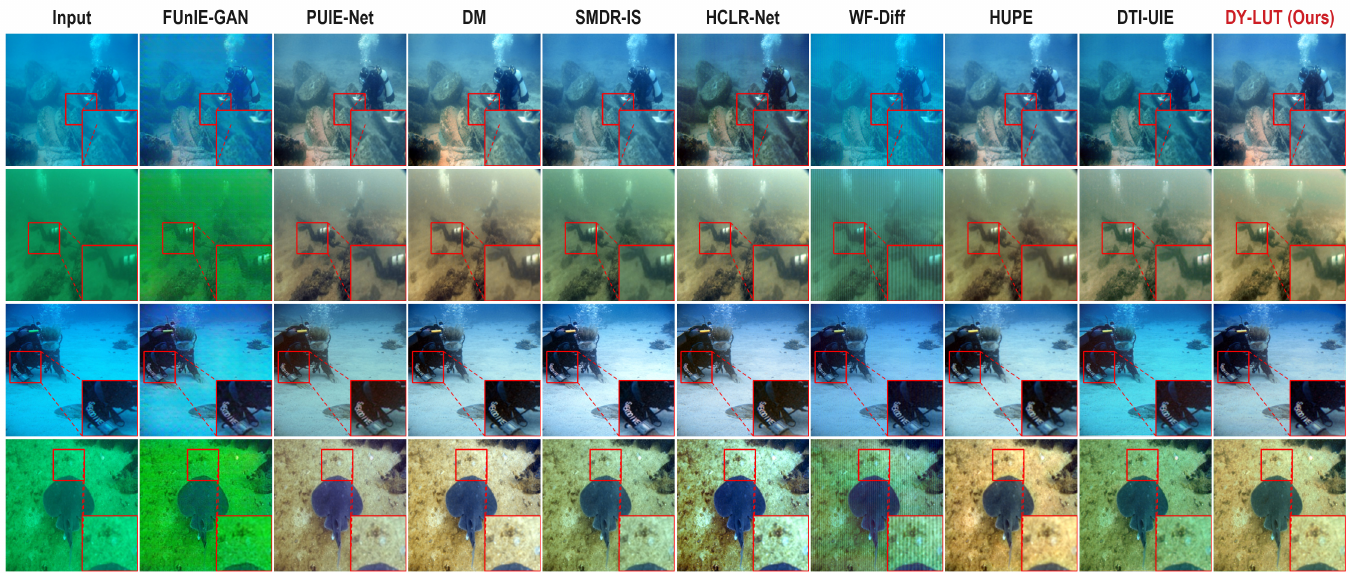}
    \caption{Qualitative comparisons on the zero-shot U45 and UIEB-C60 datasets.}
    \Description{Qualitative comparison of underwater image enhancement methods on representative images from the U45 and UIEB-C60 datasets.}
    \label{fig:generalization_qualitative}
\end{figure*}

\begin{table*}[t]
\centering
\small
\setlength{\tabcolsep}{1.15mm}
\begin{tabular}{@{}llcccccccc@{}}
\toprule
\multirow{2}{*}{\textbf{Method}} & \multirow{2}{*}{\textbf{Venue}} 
& \multicolumn{4}{c}{\textbf{U45}} 
& \multicolumn{4}{c}{\textbf{UIEB-C60}} \\
\cmidrule(lr){3-6} \cmidrule(l){7-10}
& & \textbf{UIQM}$\uparrow$ & \textbf{UCIQE}$\uparrow$ & \textbf{UISM}$\uparrow$ & \textbf{Time (ms)}$\downarrow$
  & \textbf{UIQM}$\uparrow$ & \textbf{UCIQE}$\uparrow$ & \textbf{UISM}$\uparrow$ & \textbf{Time (ms)}$\downarrow$ \\
\midrule
FUnIE-GAN & RA-L'20
& 3.3350 & 0.3999 & 8.3835 & \second{7.05}
& 3.4439 & \best{0.4418} & 8.0993 & \best{11.41} \\

PUIE-Net & ECCV'22
& \best{3.7949} & 0.3332 & 8.8028 & 80.91
& \best{4.1742} & 0.3207 & \second{10.1806} & 655.10 \\

DM & MM'23
& 3.6746 & 0.3973 & 8.9594 & 130.34
& 4.0724 & 0.3748 & 9.9789 & 1442.00 \\

SMDR-IS & AAAI'24
& 3.6896 & 0.3815 & 8.9301 & \third{33.95}
& 4.0759 & \third{0.3784} & 10.0751 & 82.81 \\

HCLR-Net & IJCV'24
& 3.6670 & \third{0.4031} & \third{9.0038} & 38.40
& 3.9643 & \second{0.3825} & 9.7173 & 275.82 \\

WF-Diff & CVPR'24
& 3.2720 & 0.3766 & 7.5530 & 278.83
& 3.9148 & 0.3748 & 9.5146 & 2025.48 \\

HUPE & IJCV'25
& 3.6317 & \best{0.4153} & 8.9560 & 103.92
& 3.9274 & 0.3661 & 9.4241 & 103.60 \\

DTI-UIE & TIP'26
& \third{3.6936} & 0.3818 & \second{9.0887} & 44.35
& \second{4.1354} & 0.3763 & \best{10.3568} & \third{44.35} \\

\textbf{DY-LUT (Ours)} & None
& \second{3.7249} & \second{0.4096} & \best{9.0952} & \best{6.65}
& \third{4.1336} & 0.3651 & \third{10.0834} & \second{12.42} \\
\bottomrule
\end{tabular}
\caption{Zero-shot comparison on U45 and UIEB-C60; ranking styles follow Tab.~\ref{tab:main_results}.}
\label{tab:generalization}
\end{table*}

\section{Experiments}

\subsection{Experimental Settings}

\textbf{Datasets.}
DY-LUT is trained only on UIEB-800~\cite{li2019underwater}. We use UIEB-90 and LSUI~\cite{peng2023u} for full-reference evaluation, U45~\cite{li2019fusion} and UIEB-C60~\cite{li2019underwater} for zero-shot no-reference evaluation, UIQAD~\cite{chu2023sisc} for 1080P/4K efficiency, RUOD~\cite{fu2023rethinking} for detection, and FLSea~\cite{randall2023flsea} for SIFT feature matching.


\textbf{Compared methods.}
We compare representative physical and learned methods: IBLA~\cite{peng2017underwater}, Sea-Thru~\cite{akkaynak2019sea}, FUnIE-GAN~\cite{islam2020fast}, PUIE-Net~\cite{fu2022uncertainty}, DM~\cite{tang2023underwater}, SMDR-IS~\cite{zhang2024synergistic}, HCLR-Net~\cite{zhou2024hclr}, WF-Diff~\cite{zhao2024wavelet}, HUPE~\cite{zhang2025hupe}, and DTI-UIE~\cite{lin2026dtiuie}. High-resolution experiments use the subset of applicable fast baselines.

\textbf{Evaluation metrics.}
Following WF-Diff~\cite{zhao2024wavelet}, we compute PSNR, SSIM~\cite{wang2004image}, and LPIPS~\cite{zhang2018unreasonable} at \(256\times256\) on paired datasets. We use UIQM, UISM~\cite{panetta2015human}, and UCIQE~\cite{yang2015underwater} otherwise. \mbox{Efficiency reports} parameters, FLOPs, latency, and FPS at each dataset's native resolution under its stated deployment.

\textbf{Implementation details.} We use AdamW~\cite{loshchilov2019decoupled} with weight decay \(10^{-5}\), learning rates \(2\times10^{-4}\) for LUTs and \(5\times10^{-4}\) for the remaining modules, and train for 800 epochs with batch size 4 and \(256\times256\) crops on one RTX 4090D. On UIQAD only, adaptive inference uses \(0.5\times\) resolution for 1080P--sub-4K inputs and \(0.25\times\) for 4K or above, followed by upsampling. We also report full-resolution results for direct comparison.

\subsection{Comparisons with SOTA Methods}

\noindent\textbf{Full-reference benchmarks.} On UIEB-90 (Tab.~\ref{tab:main_results}), DY-LUT ranks second with 23.245\,dB PSNR and 0.8994 SSIM while using only 3.56M parameters and 10.79\,ms. Only SMDR-IS obtains a higher PSNR, yet DY-LUT is $9.9\times$ faster and uses about $3.5\times$ fewer parameters. It also exceeds HCLR-Net in PSNR at a $34.3\times$ speedup and outperforms WF-Diff in PSNR while being $304\times$ faster. On zero-shot LSUI, \mbox{DY-LUT} achieves the best PSNR/SSIM and second-best LPIPS under the same training protocol. Fig.~\ref{fig:qualitative} further shows natural color recovery and spatial consistency across varied scene depths.

\noindent\textbf{Generalization on U45 and UIEB-C60.} On U45, DY-LUT ranks second in UIQM and first in UISM; on UIEB-C60, it ranks third in both metrics while retaining 6.65/12.42\,ms latency (Tab.~\ref{tab:generalization}). DTI-UIE and PUIE-Net obtain higher UIQM/UISM on UIEB-C60 but are $3.6\times$ and $52.7\times$ slower, respectively, while HCLR-Net's higher UCIQE comes at $22.2\times$ the latency. Qualitative comparisons in Fig.~\ref{fig:generalization_qualitative} further show natural color restoration under zero-shot distribution shifts. Overall, DY-LUT maintains a competitive \mbox{quality--efficiency} trade-off under unseen distribution shifts.

\subsection{Depth Source, Robustness, and Cost}
\label{sec:DepthAnalysis}

\noindent\textbf{Depth source.}
Practical systems may obtain depth directly from onboard ranging or stereo. Because public UIE benchmarks lack synchronized depth, we use DAv2-Small~\cite{yang2024depthv2} as a reproducible monocular substitute. We additionally test MiDaS v2.1-Small~\cite{ranftl2022towards}, \mbox{Lite-Mono}~\cite{zhang2023litemono}, and the classical single-image priors used by IBLA and HUPE. All inputs are normalized identically before entering the DY-LUT encoder.

\begin{center}
\begin{minipage}{\columnwidth}
\centering
\scriptsize
\setlength{\tabcolsep}{0.6mm}
\resizebox{\columnwidth}{!}{%
\begin{tabular}{@{}llccccccc@{}}
\toprule
\multirow{2}{*}{\textbf{Depth Source}}
& \multirow{2}{*}{\textbf{Arch.}}
& \multirow{2}{*}{\textbf{Params}$\downarrow$}
& \multirow{2}{*}{\textbf{FLOPs}$\downarrow$}
& \multirow{2}{*}{\textbf{Time}$\downarrow$}
& \multicolumn{2}{c}{\textbf{No Retraining}}
& \multicolumn{2}{c}{\textbf{Retrained}} \\
\cmidrule(lr){6-7}\cmidrule(l){8-9}
& & & & &
\textbf{PSNR}$\uparrow$ &
\textbf{SSIM}$\uparrow$ &
\textbf{PSNR}$\uparrow$ &
\textbf{SSIM}$\uparrow$ \\
\midrule
DAv2-Small (ref.)  & ViT         & 24.8M & 41.3\,G & 33.6\,ms & 23.245 & 0.8994 & --     & --     \\
MiDaS v2.1-Small   & CNN         & 21.3M & 4.6\,G  & 9.6\,ms  & 23.130 & 0.8970 & 23.235 & 0.8937 \\
Lite-Mono          & CNN         & 2.85M & 5.0\,G  & 7.2\,ms  & 22.900 & 0.8943 & 23.496 & 0.9047 \\
\midrule
IBLA (blurriness)  & Physical    & --    & 3.3\,G  & 3.1\,ms  & 22.827 & 0.8913 & 23.768 & 0.9004 \\
HUPE (GDCP)        & Physical    & --    & 0.01\,G & 1.1\,ms  & 22.482 & 0.8822 & 23.538 & 0.8946 \\
\bottomrule
\end{tabular}%
}
\captionof{table}{Depth-source robustness and cost on UIEB-90. Source costs are measured at native resolution and exclude DY-LUT enhancement.}
\label{tab:depth}
\end{minipage}
\end{center}

\begin{table*}[!t]
\centering
\small
\setlength{\tabcolsep}{2.0mm}
\begin{tabular}{@{}lcccccccc@{}}
\toprule
\multirow{2}{*}{\textbf{Method}}
& \multicolumn{4}{c}{\textbf{1080P}}
& \multicolumn{4}{c}{\textbf{4K}} \\
\cmidrule(lr){2-5} \cmidrule(l){6-9}
& \textbf{Time (ms)}$\downarrow$ & \textbf{FPS}$\uparrow$ & \textbf{UIQM}$\uparrow$ & \textbf{UISM}$\uparrow$
& \textbf{Time (ms)}$\downarrow$ & \textbf{FPS}$\uparrow$ & \textbf{UIQM}$\uparrow$ & \textbf{UISM}$\uparrow$ \\
\midrule
FUnIE-GAN                   & 16.97  & 58.9  & 3.1861 & 7.6374 & 57.34  & 17.4  & 3.5161 & 7.9242 \\
SMDR-IS                     & 183.37 & 5.5   & 4.4341 & 10.8098 & 726.86 & 1.4   & 4.1166 & 9.8768 \\
HUPE                        & 99.50  & 10.1  & 3.8042 & 8.8547 & 522.37  & 1.91  & 3.7012 & 8.4005 \\
DTI-UIE                     & 44.35  & 22.5  & 3.2965 & 7.0259 & 254.13  & 3.93  & 3.4222 & 7.6490 \\
\textbf{DY-LUT (Full Res.)} & 32.83  & 30.5   & 4.3325 & 10.5307 & 145.24 & 6.9   & 4.1119 & 9.7291 \\
\textbf{DY-LUT (Adaptive)}  & 7.07   & 141.4 & 4.0529 & 9.5609 & 7.08   & 141.3 & 3.7759 & 8.5495 \\
\bottomrule
\end{tabular}
\caption{Comparison at 1080P and 4K.}
\label{tab:highres_efficiency}
\end{table*}

\noindent\textbf{Necessity and robustness.}
The coordinate ablation in Tab.~\ref{tab:ablation_rep} shows consistent gains from adding depth: RGB improves from 19.10 to 20.79\,dB, while YCbCr improves from 20.87 to 21.73\,dB. Replacing the fixed chrominance coordinates with learned indices further raises the depth-conditioned YCbCr result to 23.245\,dB. Tab.~\ref{tab:depth} then separates two questions. The fixed setting changes only the depth source while retaining the DAv2-trained DY-LUT, directly measuring cross-source sensitivity; the PSNR drop remains within 0.763\,dB across all alternatives. The retrained setting independently optimizes DY-LUT for each source, with every alternative recovering to at least 23.235\,dB. Thus, the framework is not tied to one estimator and can adapt to source-specific depth characteristics.

\noindent\textbf{End-to-end cost.}
Tab.~\ref{tab:main_results} reports enhancement latency separately from optional monocular estimation. Sensor- or stereo-equipped platforms provide depth with the image and therefore incur no model-side estimation cost. When such depth is unavailable, a lightweight monocular estimator can be used: serially combining Lite-Mono (7.2\,ms) with DY-LUT (10.79\,ms) gives about 18.0\,ms, or 56 FPS. Reporting the two costs separately therefore covers both deployment settings while confirming real-time end-to-end operation.

\subsection{High-Resolution Efficiency}
\label{sec:High-ResolutionEfficiency}

On UIQAD (Tab.~\ref{tab:highres_efficiency}), adaptive inference is $4.6\times$/$20.5\times$ faster than full-resolution DY-LUT at 1080P/4K and exceeds 141 FPS on an RTX 4090D. Despite its reduced internal resolution, it surpasses FUnIE-GAN, HUPE, and DTI-UIE in both UIQM and UISM at both resolutions; only SMDR-IS scores higher, but runs at just 1.4 FPS at 4K, nearly $100\times$ slower. At 4K, DY-LUT is also over $8\times$ faster than FUnIE-GAN while preserving scene structure and natural color (Fig.~\ref{fig:4k_visual}). Its nearly identical 1080P/4K latency (7.07/7.08\,ms) confirms that downsampling bounds the internal workload. Relative to full-resolution inference, it trades 0.336 UIQM and 1.180 UISM for a $20.5\times$ speedup at 4K. Full-resolution inference remains available when maximum enhancement scores are preferred; additional native-size visual results appear in the supplement.

\noindent\textbf{Cross-device FP32 deployment.}
We further evaluate deployment across consumer, data-center, edge, and CPU hardware. All measurements use FP32, batch size one, precomputed depth, and adaptive inference; depth estimation and disk I/O are excluded. As shown in Tab.~\ref{tab:cross_device}, the bounded internal resolution keeps 1080P and 4K throughput close on every device. The Tesla T4 is fastest among these additional platforms at 19.02/18.23 FPS, while the Jetson Orin NX reaches 10.39/10.59 FPS. Even the CPU-only configuration remains above one FPS at both resolutions, demonstrating that DY-LUT can operate across a broad range of deployment targets.

\begin{table}[H]
\centering
\small
\setlength{\tabcolsep}{8pt}
\renewcommand{\arraystretch}{1.08}
\begin{tabular}{@{}llrr@{}}
\toprule
\textbf{Hardware} & \textbf{Type} & \textbf{1080P FPS} & \textbf{4K FPS} \\
\midrule
AMD Ryzen 7 4800H        & CPU             &  1.24 &  1.10 \\
GeForce GTX 750 Ti       & Desktop GPU     &  4.46 &  4.46 \\
Jetson Orin NX 8GB       & Edge GPU        & 10.39 & 10.59 \\
Tesla P4                 & Data-center GPU & 14.32 & 14.31 \\
GeForce RTX 2060         & Desktop GPU     &  9.64 &  8.51 \\
Tesla T4                 & Data-center GPU & \textbf{19.02} & \textbf{18.23} \\
\bottomrule
\end{tabular}
\caption{Cross-device FP32 network throughput in FPS (higher is better).}
\label{tab:cross_device}
\end{table}

\begin{figure}[!t]
\begin{minipage}{\linewidth}
    \centering
    \includegraphics[width=0.90\linewidth]{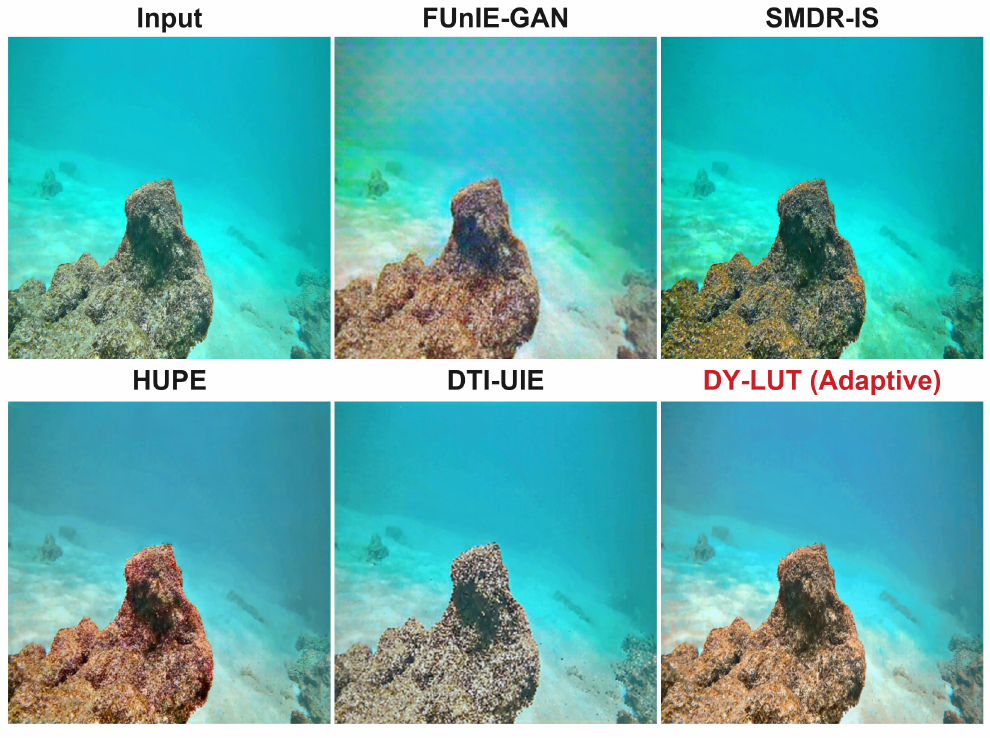}
    \caption{4K qualitative comparison on UIQAD; DY-LUT uses adaptive inference.}
    \Description{A 4K UIQAD scene enhanced by FUnIE-GAN, SMDR-IS, HUPE, DTI-UIE, and adaptive DY-LUT. DY-LUT restores the rock color while preserving the surrounding water and background structure.}
    \label{fig:4k_visual}
\centering
\small
\setlength{\tabcolsep}{0.20mm}
\renewcommand{\arraystretch}{1.00}
\begin{tabular}{@{}lcccc@{}}
\toprule
\multirow{2}{*}{\textbf{Method}}
& \multicolumn{2}{c}{\textbf{RUOD Detection}}
& \multicolumn{2}{c}{\textbf{FLSea Matching}} \\
\cmidrule(lr){2-3}\cmidrule(l){4-5}
& \textbf{mAP@0.5}$\uparrow$ & \textbf{mAP@0.5:0.95}$\uparrow$
& \textbf{Matches}$\uparrow$ & \textbf{Inliers}$\uparrow$ \\
\midrule
FUnIE-GAN              & 0.8131 & 0.5964 & 450 & 339 \\
SMDR-IS                & 0.8399 & 0.6243 & 612 & 446 \\
HUPE                   & 0.8045 & 0.5954 & 462 & 325 \\
DTI-UIE                & 0.8436 & 0.6297 & 535 & \textbf{476} \\
\textbf{DY-LUT (Ours)} & \textbf{0.8578} & \textbf{0.6340} & \textbf{661} & 470 \\
\bottomrule
\end{tabular}
\captionof{table}{Downstream results on RUOD and FLSea.}
\label{tab:downstream}
\end{minipage}
\end{figure}

\subsection{Downstream Tasks}
\label{sec:DownstreamPerception}

For detection, we fix a YOLO11n model~\cite{jocher2024yolo11} fine-tuned on RUOD, so only the preceding enhancement method changes. For matching, we use OpenCV SIFT~\cite{lowe2004distinctive,Bradski2000OpenCV} on FLSea and report mean match and inlier counts. As shown in Tab.~\ref{tab:downstream}, DY-LUT obtains the best detection accuracy and the largest number of matches. Against DTI-UIE, the strongest common detection baseline, DY-LUT improves mAP@0.5 by 1.42 percentage points and mAP@0.5:0.95 by 0.43 points. It also produces 126 more SIFT matches, while the six-inlier deficit indicates that the additional correspondences remain largely geometrically consistent. The agreement between restoration and task-level results shows that efficiency is not obtained by discarding structures needed by later vision modules. Qualitative visualizations are provided in the supplement.

\subsection{Ablation Studies}

All ablations follow the UIEB-90 full-reference protocol.
\noindent\textbf{Representation ablation.}
Tab.~\ref{tab:ablation_rep} compares direct RGB and YCbCr
lookup coordinates, their depth-augmented variants, and the
proposed learned-coordinate design. RGB and YCbCr use direct
color coordinates, while RGBD and YCbCrD additionally use the
normalized depth coordinate \(D\). DY-LUT replaces the fixed
chrominance coordinates \((C_b,C_r)\) in YCbCrD with the jointly
learned latent coordinates
\((\CDIone,\CDItwo)\). Each variant is trained independently.
Adding depth improves PSNR from 19.10 to 20.79\,dB in RGB and
from 20.87 to 21.73\,dB in YCbCr. YCbCr outperforms RGB in both
settings, while learned coordinates further improve the
depth-conditioned YCbCr design to 23.245\,dB.

\begin{table}[!htbp]
\centering
\small
\renewcommand{\arraystretch}{1.04}
\begin{tabular*}{\columnwidth}{@{\extracolsep{\fill}}lccc@{}}
\toprule
\textbf{Method} &
\textbf{Coordinates} &
\textbf{PSNR}$\uparrow$ &
\textbf{SSIM}$\uparrow$ \\
\midrule
RGB LUT
& $(R,G,B)$
& 19.10
& 0.8370 \\

RGBD LUT
& $(R,G,B,D)$
& 20.79
& 0.8558 \\

YCbCr LUT
& $(Y,C_b,C_r)$
& 20.87
& 0.8741 \\

YCbCrD LUT
& $(Y,D,C_b,C_r)$
& 21.73
& 0.8769 \\

\textbf{DY-LUT (Ours)}
& $(Y,D,\CDIone,\CDItwo)$
& \textbf{23.245}
& \textbf{0.8994} \\
\bottomrule
\end{tabular*}
\caption{Representation and lookup-coordinate ablation on UIEB-90. Coordinates list the LUT axes; \(D\) is normalized depth and \(\CDIone,\CDItwo\) are learned latent coordinates.}
\label{tab:ablation_rep}
\end{table}

\begin{table}[!htbp]
\centering
\small
\setlength{\tabcolsep}{0.45mm}
\renewcommand{\arraystretch}{1.08}
\begin{tabular}{@{}llccc@{}}
\toprule
\textbf{Group} & \textbf{Variant} & \textbf{Removed} & \textbf{PSNR}$\uparrow$ & \textbf{SSIM}$\uparrow$ \\
\midrule
\multirow{3}{*}{Arch.}
& w/o Weight Branch & 1.2K & 20.37 & 0.8793 \\
& w/o Local Refine  & 1.7K & 21.07 & 0.8826 \\
& w/o Gradient Cue  & 0.3K & 22.03 & 0.8901 \\
\midrule
\multirow{3}{*}{Loss}
& w/o $\mathcal{L}_{\mathrm{MN}}$   & -- & 21.84 & 0.8919 \\
& w/o $\mathcal{L}_{\mathrm{VGG}}$  & -- & 21.75 & 0.8750 \\
& w/o $\mathcal{L}_{\mathrm{grad}}$ & -- & 22.05 & 0.8894 \\
\midrule
\multicolumn{2}{@{}l}{\textbf{DY-LUT (Ours)}} & -- & \textbf{23.245} & \textbf{0.8994} \\
\bottomrule
\end{tabular}
\caption{Architecture and loss ablations with parameter costs.}
\label{tab:ablation_arch_loss}
\end{table}

\noindent\textbf{Architectural ablation.} Removing the Weight Branch produces the largest drop, from 23.245 to 20.37\,dB, confirming that image-level LUT fusion is central (Tab.~\ref{tab:ablation_arch_loss}). Removing Local Refinement or the gradient cue yields 21.07 and 22.03\,dB, supporting their complementary spatial and structural roles. The three modules add only 1.2K, 1.7K, and 0.3K parameters; the supplement reports LUT-capacity sweeps.

\noindent\textbf{Loss ablation.} Removing $\mathcal{L}_{\mathrm{MN}}$, $\mathcal{L}_{\mathrm{VGG}}$, or $\mathcal{L}_{\mathrm{grad}}$ gives 21.84, 21.75, and 22.05\,dB, respectively (Tab.~\ref{tab:ablation_arch_loss}); the VGG-free variant also has the lowest SSIM. The results support complementary roles for monotonic LUT ordering, perceptual appearance, and edge preservation.

\noindent\textbf{Joint degradation-index representation.} Replacing the fixed chrominance coordinates \((C_b,C_r)\) in YCbCrD LUT with the jointly learned pair \((\CDIone,\CDItwo)\) improves PSNR from 21.73 to 23.245\,dB and SSIM from 0.8769 to 0.8994 (Tab.~\ref{tab:ablation_rep}). This result shows that content-adaptive latent coordinates are more effective than direct chrominance coordinates for depth-conditioned lookup. Because the learned indices are non-identifiable, we interpret them only as a joint representation rather than assigning separate physical meanings to the two channels.

\begin{figure}[!htbp]
    \centering
    \includegraphics[width=0.96\linewidth]{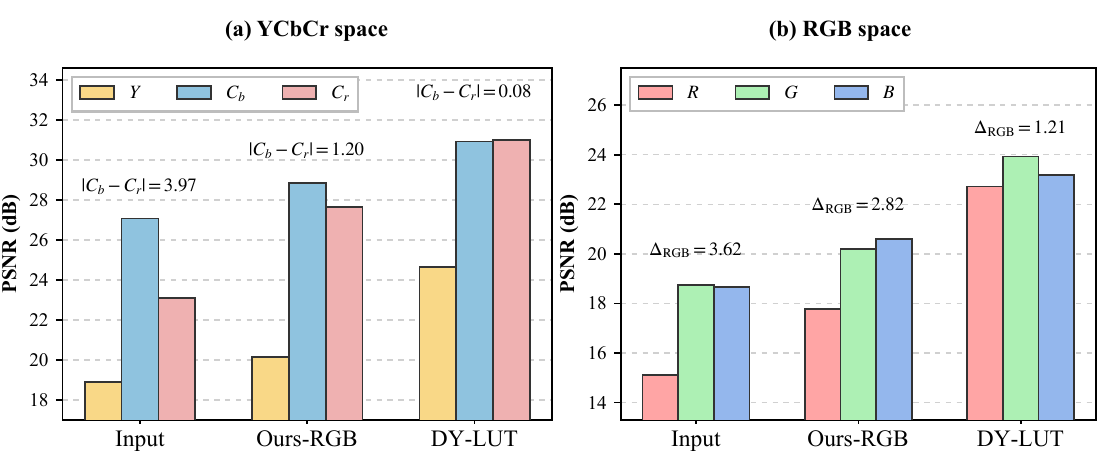}
    \caption{Channel-wise PSNR analysis in the YCbCr and RGB spaces on UIEB-90.}
    \Description{Channel-wise PSNR comparison of the input, Ours-RGB, and DY-LUT. The left panel shows $Y$, $C_b$, and $C_r$ together with the absolute $C_b$--$C_r$ discrepancy, while the right panel shows $R$, $G$, and $B$ together with the RGB channel discrepancy.}
    
    \label{fig:channel_psnr}
\end{figure}

\noindent\textbf{Merits of YCbCr space.} In Fig.~\ref{fig:channel_psnr}, the $C_b$--$C_r$ PSNR gap falls from 3.97 for the input and 1.20 for Ours-RGB to 0.08 for DY-LUT. After RGB conversion, DY-LUT's channel discrepancy is 1.21 versus 2.82 for Ours-RGB. Because the two variants retain the same depth-conditioned, learned-index design and differ in their operating color space, the more balanced chrominance and red-channel recovery supports YCbCr as the more effective restoration basis.

\noindent\textbf{Discussion.} The lookup-coordinate ablation shows that YCbCr coordinates outperform RGB coordinates in both the no-depth and depth-conditioned settings, while adding depth improves both direct-color representations. Replacing fixed \((C_b,C_r)\) coordinates with the learned index pair further improves the depth-conditioned YCbCr design. The learned channels are non-identifiable, so we interpret only their joint representation. Architecture and loss ablations provide complementary evidence. Unlike analytical inversion, DY-LUT conditions a learned residual on depth rather than imposing a fixed transmission law, supporting adaptation across depth sources without changing the lookup mechanism.

\noindent\textbf{Limitations.} Enhancement quality depends on the reliability of the supplied depth cue, particularly in textureless or highly turbid regions. Adaptive downsampling can also suppress fine details at 4K, motivating confidence-aware depth conditioning and content-adaptive resolution selection.

\section{Conclusion}

We present DY-LUT, a depth-aware YCbCr 4D LUT framework for efficient underwater enhancement. Its dual-branch encoder predicts image-level fusion weights and latent lookup coordinates, while local refinement preserves spatial consistency. With supplied depth, the 3.56M-parameter enhancement network runs in 10.79\,ms on UIEB-90 and exceeds 141 FPS with adaptive 4K inference. Paired, zero-shot, high-resolution, and downstream evaluations demonstrate a favorable quality--efficiency balance. Platform sensors or lightweight monocular estimation support sensor-assisted and RGB-only deployment. Unlike iterative models, DY-LUT retains a fixed feed-forward computational path while its LUT weights and query coordinates remain content-adaptive, providing predictable enhancement without a dense image-to-image backbone.

More broadly, scene geometry organizes a compact restoration space with transparent computation and predictable scaling on resource-constrained platforms. Its modular depth interface supports varied sensors without redesigning the enhancement pipeline.

\clearpage
\appendix
\section*{Supplementary Material}
\addcontentsline{toc}{section}{Supplementary Material}

\section{Overview}

This supplement expands the implementation and visual evidence of the main paper. It provides the exact quadrilinear lookup, detailed inference and training pseudocode, LUT-capacity scaling, dataset-wise qualitative comparisons, native-size 1080P and 4K examples, and downstream detection and feature-matching visualizations. Unless stated otherwise, all checkpoints, datasets, and evaluation settings follow Sec.~4.1 of the main paper. DY-LUT is trained only on UIEB-800~\cite{li2019underwater}; LSUI, U45, UIEB-C60, and UIQAD are evaluated without additional fine-tuning.

\section{Exact 4D LUT Evaluation}

DY-LUT uses \(K=3\) learnable tables
\(\mathcal{T}_k\in\mathbb{R}^{N\times N\times N\times N\times3}\)
with \(N=25\). For a normalized per-pixel query
\begin{equation}
\mathbf{q}=[q_1,q_2,q_3,q_4]
=[Y,D,\CDIone,\CDItwo]\in[0,1]^4,
\end{equation}
we compute \(\mathbf{u}=(N-1)\mathbf{q}\). For each axis \(j\), the lower vertex and fractional offset are
\begin{equation}
i_j=\min\!\left(\lfloor u_j\rfloor,N-2\right),
\qquad \delta_j=u_j-i_j.
\end{equation}
The quadrilinear output of table \(\mathcal{T}_k\) is
\begin{equation}
\label{eq:supp_quad}
\begin{aligned}
w(\mathbf{b})
&=\prod_{j=1}^{4}\bigl(b_j\delta_j+(1-b_j)(1-\delta_j)\bigr),\\
\operatorname{Interp}(\mathcal{T}_k;\mathbf{q})
&=\sum_{\mathbf{b}\in\{0,1\}^{4}}
w(\mathbf{b})\,\mathcal{T}_k[\mathbf{i}+\mathbf{b}].
\end{aligned}
\end{equation}
Thus, each query combines 16 adjacent vertices with separable linear weights. Clamping \(\mathbf{q}\) and limiting the lower index to \(N-2\) make the interpolation valid at both boundaries. The final residual is the image-adaptive mixture \(\sum_{k=1}^{K}\alpha_k\operatorname{Interp}(\mathcal{T}_k;\mathbf{q})\). The two learned channels \([\CDIone,\CDItwo]\) are optimized jointly as latent lookup coordinates; the method does not assign either channel a separate physical interpretation.

\section{Detailed Pseudocode}

Algorithms~\ref{alg:supp_inference} and~\ref{alg:supp_training} use the same notation as the main paper. In particular, \(\boldsymbol{\alpha}\) denotes image-level LUT weights, \(\mathbf{C}_{\mathrm{DI}}=[\CDIone,\CDItwo]\) denotes the joint pixel-wise index field, and \(\mathcal{T}_k\) denotes the \(k\)-th 4D LUT. This avoids introducing duplicate symbols for the same quantities.

\begin{algorithm}[H]
\small
\caption{DY-LUT Inference}
\label{alg:supp_inference}
\begin{algorithmic}[1]
\REQUIRE RGB image \(I_{\mathrm{RGB}}\in[0,1]^{3\times H\times W}\)
\REQUIRE Normalized depth \(D\in[0,1]^{1\times H\times W}\)
\REQUIRE LUT bank \(\{\mathcal{T}_k\}_{k=1}^{K}\), shared extractor \(f_s\), prediction heads, and refinement module \(f_{\mathrm{ref}}\)
\ENSURE Enhanced image \(\hat I_{\mathrm{RGB}}\)
\STATE \textbf{Color representation and structural cue}
\STATE \(I_{\mathrm{YCbCr}}=[Y,C_b,C_r]\leftarrow\operatorname{RGB2YCbCr}(I_{\mathrm{RGB}})\)
\STATE \(\Gamma\leftarrow\operatorname{Sobel}(Y)\)
\STATE \(\mathbf{X}\leftarrow\operatorname{cat}[Y,C_b,C_r,D,\Gamma]\)
\STATE \textbf{Dual-branch degradation encoding}
\STATE \(\mathbf{F}_{s}\leftarrow f_s(\mathbf{X})\)
\STATE \(\mathbf{z}\leftarrow\operatorname{Pool}(\mathbf{F}_{s})\)
\STATE \(\boldsymbol{\alpha}\leftarrow\operatorname{Softmax}(\operatorname{Head}_{\mathrm w}(\mathbf{z}))\)
\STATE \(\mathbf{C}_{\mathrm{DI}}=[\CDIone,\CDItwo]\leftarrow\sigma(\operatorname{Head}_{\mathrm{cd}}(\mathbf{F}_{s}))\)
\STATE \textbf{Pixel-wise 4D coordinates}
\STATE \(\mathbf{q}\leftarrow\operatorname{clamp}([Y,D,\CDIone,\CDItwo],0,1)\)
\STATE \(\mathbf{u}\leftarrow(N-1)\mathbf{q}\)
\STATE \(\mathbf{i}\leftarrow\min(\lfloor\mathbf{u}\rfloor,N-2)\)
\STATE \(\boldsymbol{\delta}\leftarrow\mathbf{u}-\mathbf{i}\)
\STATE \textbf{Adaptive LUT query}
\FOR{\(k=1\) to \(K\)}
  \STATE \(R_k\leftarrow\operatorname{Interp}(\mathcal{T}_k;\mathbf{i},\boldsymbol{\delta})\) using Eq.~\eqref{eq:supp_quad}
\ENDFOR
\STATE \(\Delta I_{\mathrm{YCbCr}}^{\mathrm{LUT}}\leftarrow\sum_{k=1}^{K}\alpha_kR_k\)
\STATE \(\hat I_{\mathrm{YCbCr}}^{\mathrm{LUT}}\leftarrow I_{\mathrm{YCbCr}}+\Delta I_{\mathrm{YCbCr}}^{\mathrm{LUT}}\)
\STATE \textbf{Local refinement and output conversion}
\STATE \(\mathbf{X}_{\mathrm{ref}}\leftarrow\operatorname{cat}[\hat I_{\mathrm{YCbCr}}^{\mathrm{LUT}},I_{\mathrm{YCbCr}}]\)
\STATE \(\Delta I_{\mathrm{YCbCr}}^{\mathrm{ref}}\leftarrow f_{\mathrm{ref}}(\mathbf{X}_{\mathrm{ref}})\)
\STATE \(\hat I_{\mathrm{YCbCr}}^{\mathrm{ref}}\leftarrow\hat I_{\mathrm{YCbCr}}^{\mathrm{LUT}}+\Delta I_{\mathrm{YCbCr}}^{\mathrm{ref}}\)
\STATE \(\hat I_{\mathrm{RGB}}\leftarrow\operatorname{YCbCr2RGB}(\hat I_{\mathrm{YCbCr}}^{\mathrm{ref}})\)
\RETURN \(\hat I_{\mathrm{RGB}}\)
\end{algorithmic}
\end{algorithm}

\begin{algorithm}[H]
\small
\caption{DY-LUT Training}
\label{alg:supp_training}
\begin{algorithmic}[1]
\REQUIRE Training set \(\mathcal{D}_{\mathrm{train}}=\{(I_{\mathrm{RGB}}^{(i)},D^{(i)},I_{\mathrm{gt}}^{(i)})\}_{i=1}^{M}\)
\REQUIRE LUT bank \(\{\mathcal{T}_k\}_{k=1}^{K}\), encoder parameters, and refinement parameters
\REQUIRE Loss coefficients specified after the algorithm
\FOR{each training iteration}
  \STATE Sample a mini-batch \((I_{\mathrm{RGB}},D,I_{\mathrm{gt}})\)
  \STATE \(\hat I_{\mathrm{RGB}}\leftarrow\operatorname{DY\text{-}LUT}(I_{\mathrm{RGB}},D)\) using Algorithm~\ref{alg:supp_inference}
  \STATE \([\hat Y,\hat C_b,\hat C_r]\leftarrow\operatorname{RGB2YCbCr}(\hat I_{\mathrm{RGB}})\)
  \STATE \([Y_{\mathrm{gt}},C_{b,\mathrm{gt}},C_{r,\mathrm{gt}}]\leftarrow\operatorname{RGB2YCbCr}(I_{\mathrm{gt}})\)
  \STATE \textbf{YCbCr reconstruction objective}
  \STATE \(\mathcal{L}_{\mathrm{rec}}\leftarrow\|\hat Y-Y_{\mathrm{gt}}\|_1\)
  \STATE \(\mathcal{L}_{\mathrm{rec}}\leftarrow\mathcal{L}_{\mathrm{rec}}+\beta_{C_b}\|\hat C_b-C_{b,\mathrm{gt}}\|_1\)
  \STATE \(\mathcal{L}_{\mathrm{rec}}\leftarrow\mathcal{L}_{\mathrm{rec}}+\beta_{C_r}\|\hat C_r-C_{r,\mathrm{gt}}\|_1\)
  \STATE \(\mathcal{L}_{\mathrm{rec}}\leftarrow\mathcal{L}_{\mathrm{rec}}+\beta_{\mathrm{SSIM}}(1-\operatorname{SSIM}(\hat Y,Y_{\mathrm{gt}}))\)
  \STATE \textbf{LUT regularization and detail preservation}
  \STATE \(\mathcal{L}_{\mathrm{TV}}\leftarrow\sum_{k=1}^{K}\operatorname{TotalVariation}(\mathcal{T}_k)\)
  \STATE \(\mathcal{L}_{\mathrm{MN}}\leftarrow\sum_{k=1}^{K}\operatorname{Monotonicity}(\mathcal{T}_k;Y\text{ axis})\)
  \STATE \(\mathcal{L}_{\mathrm{VGG}}\leftarrow\operatorname{VGGPerceptual}(\hat I_{\mathrm{RGB}},I_{\mathrm{gt}})\)
  \STATE \(\mathcal{L}_{\mathrm{grad}}\leftarrow\|\nabla\hat Y-\nabla Y_{\mathrm{gt}}\|_1\)
  \STATE \textbf{Total objective and optimization}
  \STATE \(\mathcal{L}_{\mathrm{total}}\leftarrow\mathcal{L}_{\mathrm{rec}}+\beta_{\mathrm{TV}}\mathcal{L}_{\mathrm{TV}}+\beta_{\mathrm{MN}}\mathcal{L}_{\mathrm{MN}}\)
  \STATE \(\mathcal{L}_{\mathrm{total}}\leftarrow\mathcal{L}_{\mathrm{total}}+\beta_{\mathrm{VGG}}\mathcal{L}_{\mathrm{VGG}}+\beta_{\mathrm{grad}}\mathcal{L}_{\mathrm{grad}}\)
  \STATE Backpropagate \(\nabla\mathcal{L}_{\mathrm{total}}\) through the LUT bank, encoder, and refinement module
  \STATE Update all learnable parameters using AdamW~\cite{loshchilov2019decoupled}
\ENDFOR
\end{algorithmic}
\end{algorithm}

We use \(\beta_{C_b}=\beta_{C_r}=1.5\), \(\beta_{\mathrm{SSIM}}=1.0\), \(\beta_{\mathrm{TV}}=5\times10^{-5}\), \(\beta_{\mathrm{MN}}=2.0\), \(\beta_{\mathrm{VGG}}=0.1\), and \(\beta_{\mathrm{grad}}=0.05\). Gradients from all five loss components jointly update the encoder, refinement module, and LUT entries; the depth estimator is not part of this optimization.

\section{LUT Capacity Scaling}

For \(K\) tables with \(N\) bins per axis, the LUT bank contains \(3KN^4\) trainable values, while the parameter counts in Tables~\ref{tab:supp_lut_number} and~\ref{tab:supp_lut_resolution} include the lightweight encoder and refinement module. The two tables vary \(K\) and \(N\) separately while holding the other quantity fixed. With \(N=25\), performance peaks at \(K=3\), and adding more LUTs provides no further gain. Increasing the resolution to \(N=33\) improves PSNR by 0.611 dB but raises the total parameter count from 3.56M to 10.72M; still larger resolutions degrade performance despite substantially higher costs. We therefore select \((K,N)=(3,25)\) as the quality--efficiency trade-off.

\begin{table}[H]
\centering
\scriptsize
\captionsetup{skip=3pt}
\caption{Effect of the LUT bank size under a fixed LUT resolution of 25 bins.}
\label{tab:supp_lut_number}
\setlength{\tabcolsep}{3.6mm}
\renewcommand{\arraystretch}{0.94}
\begin{tabular}{@{}ccc@{}}
\toprule
\textbf{\#LUTs} & \textbf{PSNR (dB)}$\uparrow$ & \textbf{Params (M)}$\downarrow$ \\
\midrule
1 & 20.486 & 1.21 \\
2 & 21.972 & 2.39 \\
3 & 23.245 & 3.56 \\
4 & 23.183 & 4.73 \\
5 & 23.043 & 5.91 \\
\bottomrule
\end{tabular}
\end{table}

\begin{table}[H]
\centering
\scriptsize
\captionsetup{skip=3pt}
\caption{Effect of the LUT resolution under a fixed LUT bank size of 3.}
\label{tab:supp_lut_resolution}
\setlength{\tabcolsep}{3.6mm}
\renewcommand{\arraystretch}{0.94}
\begin{tabular}{@{}ccc@{}}
\toprule
\textbf{Bins} & \textbf{PSNR (dB)}$\uparrow$ & \textbf{Params (M)}$\downarrow$ \\
\midrule
17 & 21.453 & 0.81 \\
25 & 23.245 & 3.56 \\
33 & 23.856 & 10.72 \\
45 & 22.794 & 36.95 \\
65 & 22.374 & 160.71 \\
\bottomrule
\end{tabular}
\end{table}

\section{Dataset-Wise Qualitative Comparisons}

The visual comparisons cover physics-based IBLA~\cite{peng2017underwater} and Sea-Thru~\cite{akkaynak2019sea}, FUnIE-GAN~\cite{islam2020fast}, PUIE-Net~\cite{fu2022uncertainty}, DM~\cite{tang2023underwater}, SMDR-IS~\cite{zhang2024synergistic}, HCLR-Net~\cite{zhou2024hclr}, WF-Diff~\cite{zhao2024wavelet}, HUPE~\cite{zhang2025hupe}, and DTI-UIE~\cite{lin2026dtiuie}. All panels use the same source image and identical crop coordinates, so color, contrast, and local detail can be compared directly.

\subsection{UIEB-90}

\begin{figure*}[!tp]
\centering
\centering
\includegraphics[width=0.94\linewidth]{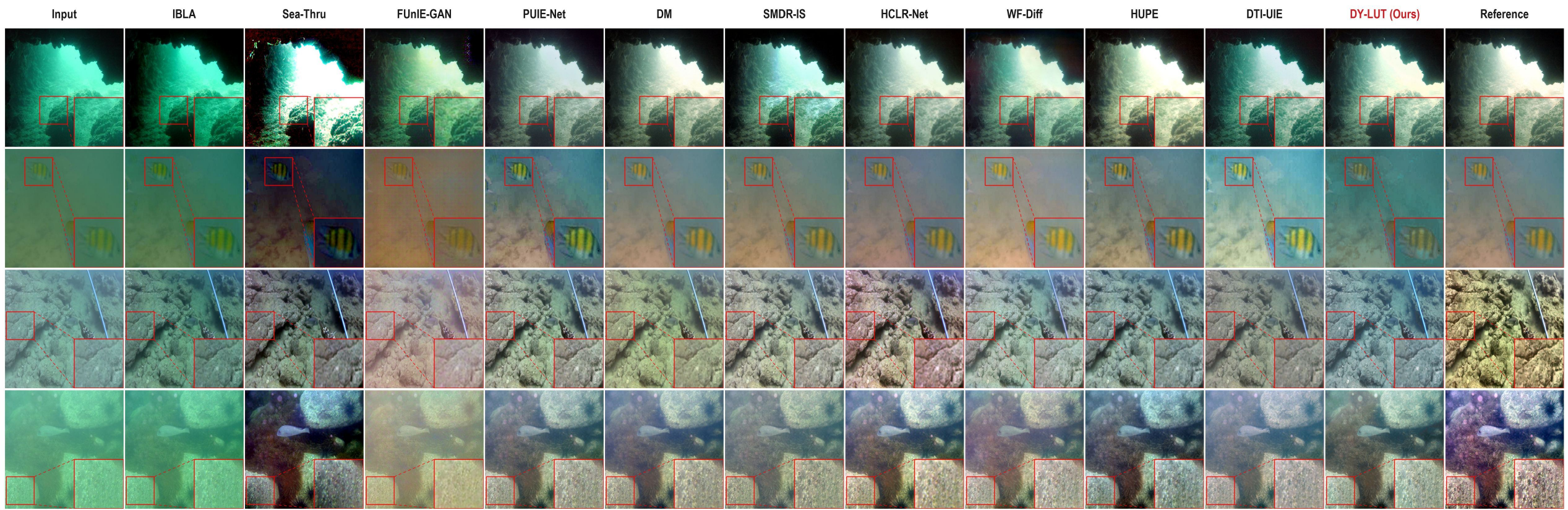}
\captionof{figure}{Qualitative comparison on UIEB-90; red boxes mark identical enlarged regions.}
\label{fig:supp_uieb90}
\end{figure*}

UIEB-90 is the in-domain paired test set. Figure~\ref{fig:supp_uieb90} covers severe illumination imbalance, green or blue casts, foreground subjects, and distant backgrounds. The reference column exposes both under-correction and over-enhancement. DY-LUT restores visibility while retaining local structures inside the shared red crops.

\subsection{LSUI}

\begin{figure*}[!tp]
\centering
\includegraphics[width=0.94\linewidth]{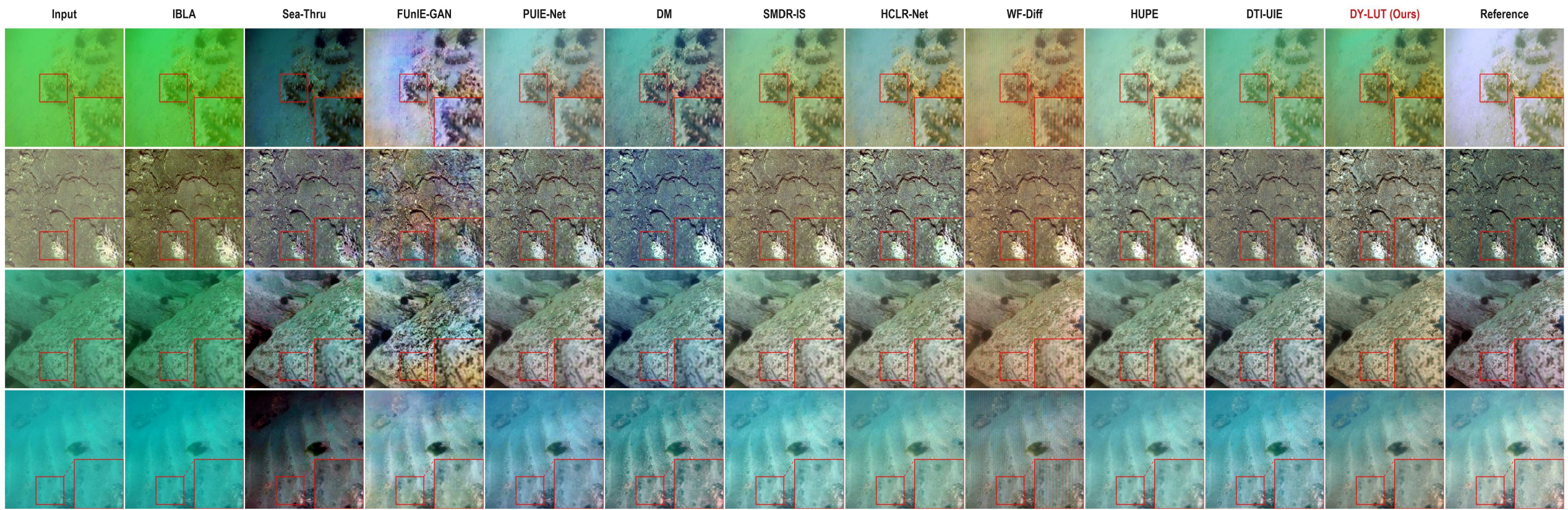}
\captionof{figure}{Zero-shot qualitative comparison on LSUI. The reference and aligned enlarged regions facilitate inspection of color fidelity and texture retention.}
\label{fig:supp_lsui}
\end{figure*}

LSUI~\cite{peng2023u} is evaluated zero-shot despite providing paired references. The scenes in Fig.~\ref{fig:supp_lsui} contain different water colors, illumination levels, and propagation distances. DY-LUT suppresses dominant casts without the excessive saturation or local contrast reversal visible in several alternatives, while preserving rocks and small foreground textures.

\subsection{U45}

\begin{figure*}[!tp]
\centering
\centering
\includegraphics[width=0.94\linewidth]{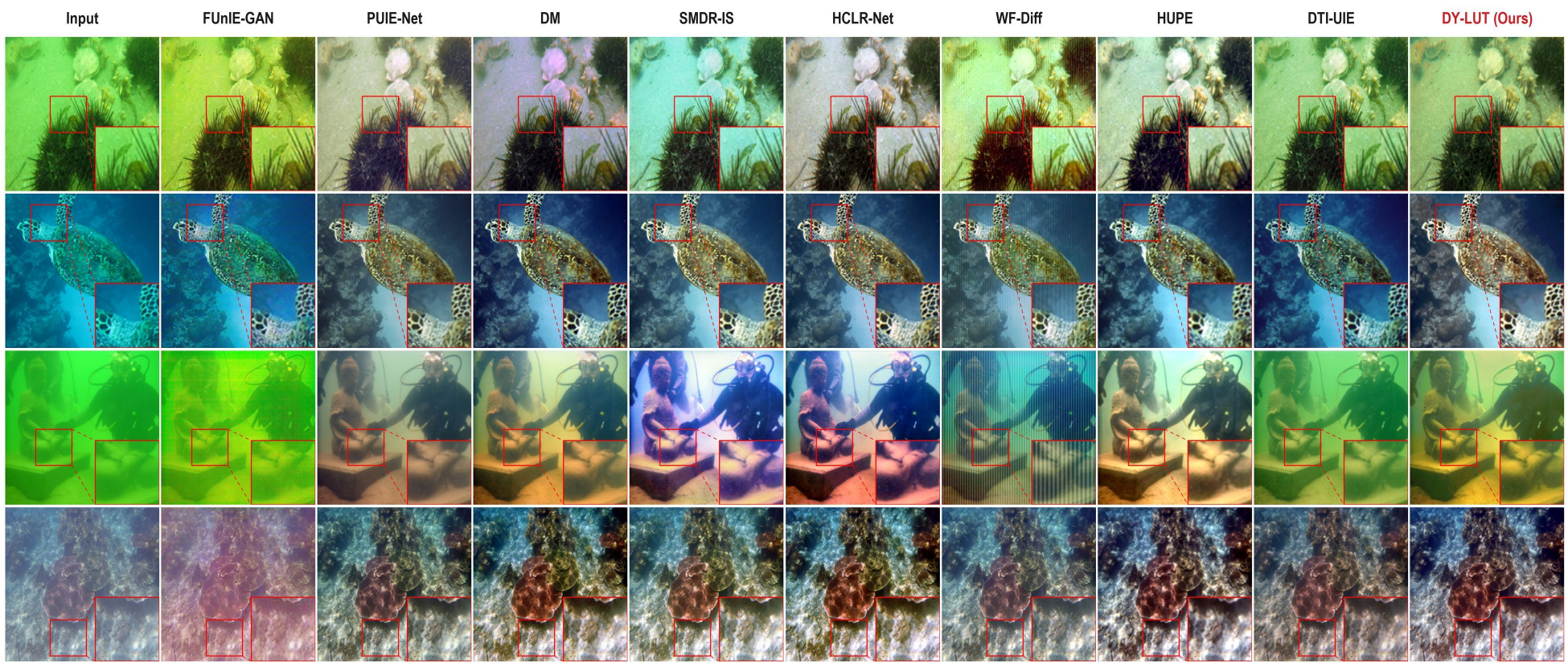}
\captionof{figure}{Zero-shot qualitative comparison on U45. The four rows cover plants, animals, people, and coral scenes under markedly different degradation conditions.}
\label{fig:supp_u45}
\end{figure*}

U45~\cite{li2019fusion} contains unpaired real underwater scenes with diverse casts and turbidity. Figure~\ref{fig:supp_u45} therefore emphasizes visible naturalness, local contrast, and structural preservation rather than agreement with a target. DY-LUT improves foreground and background visibility while limiting clipping in already bright regions.

\subsection{UIEB-C60}

\begin{figure*}[!t]
\centering
\centering
\includegraphics[width=0.94\linewidth]{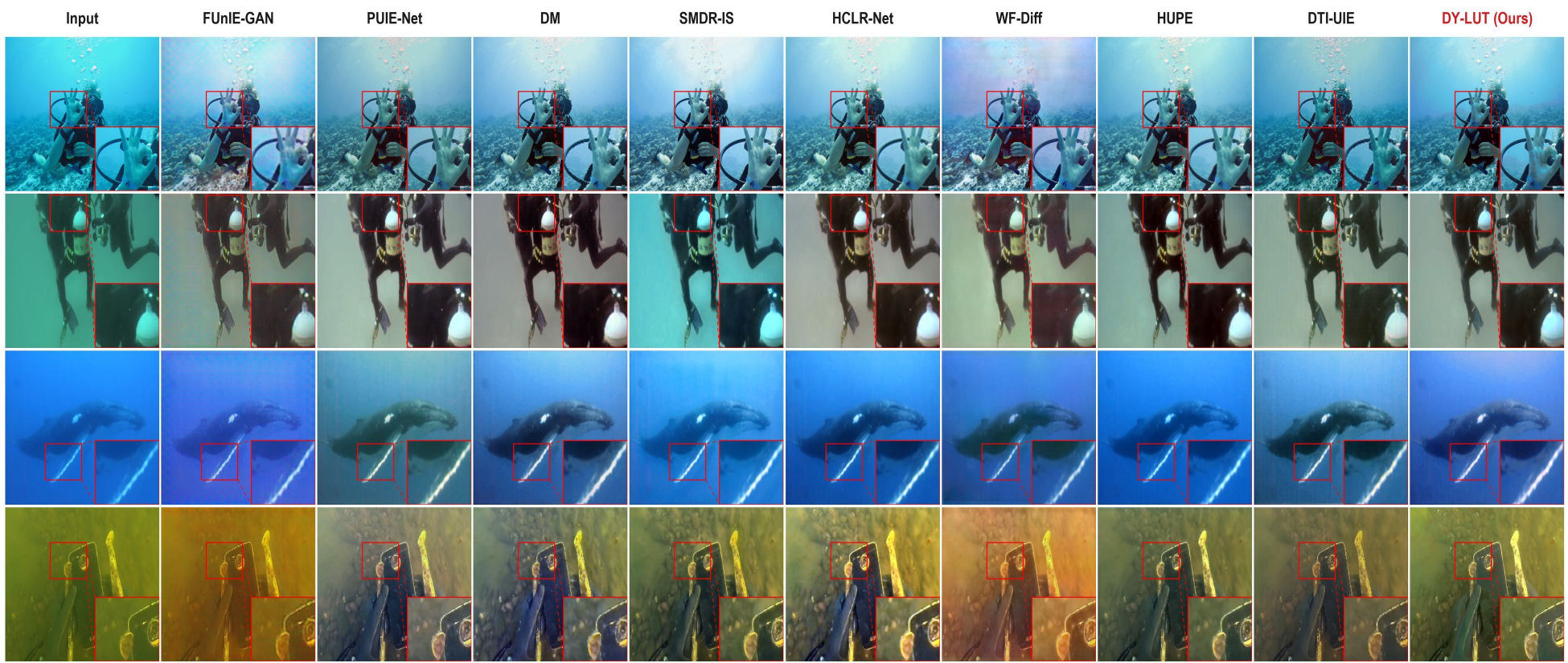}
\captionof{figure}{Zero-shot qualitative comparison on UIEB-C60. Aligned crops highlight small objects and boundaries under strong casts and spatially non-uniform attenuation.}
\label{fig:supp_c60}
\end{figure*}

UIEB-C60~\cite{li2019underwater} contains challenging images without paired references. Strong attenuation and non-uniform casts make global correction prone to residual color bias, clipping, or amplified haze. As shown in Fig.~\ref{fig:supp_c60}, depth-conditioned lookup gives DY-LUT spatial flexibility while the local module retains boundaries in divers, animals, and equipment.

\section*{High-Resolution Results}

\begin{figure}[!t]
\centering
\includegraphics[width=\columnwidth]{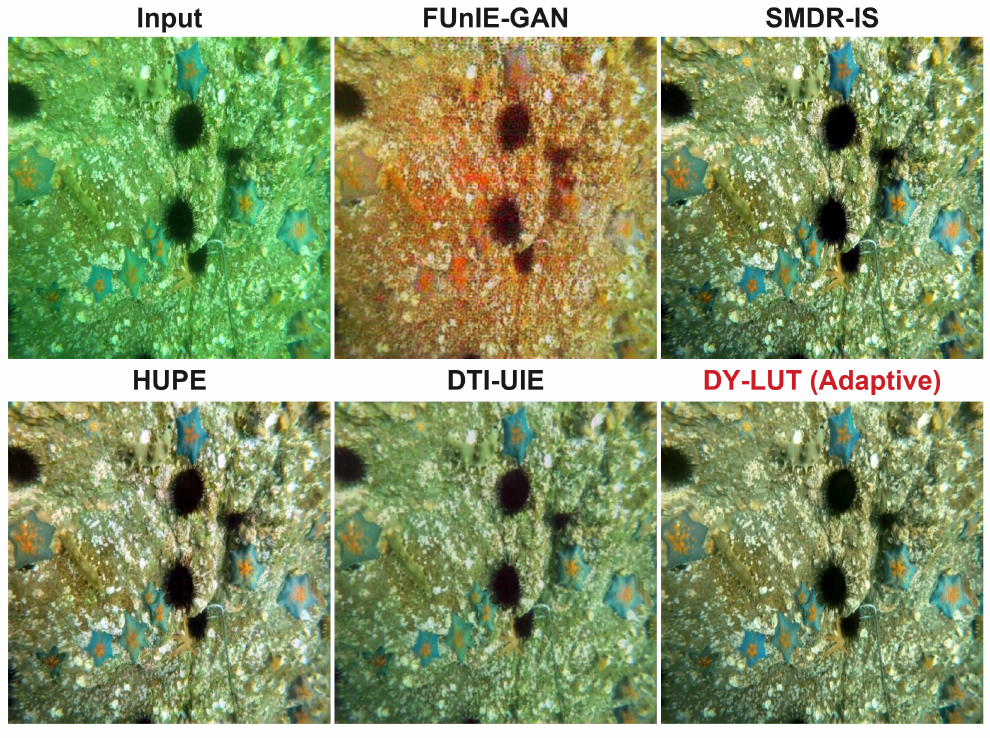}
\caption{Adaptive-inference comparison on native-size 1080P UIQAD inputs.}
\label{fig:supp_1080}
\end{figure}

\begin{figure}[!t]
\centering
\includegraphics[width=\columnwidth]{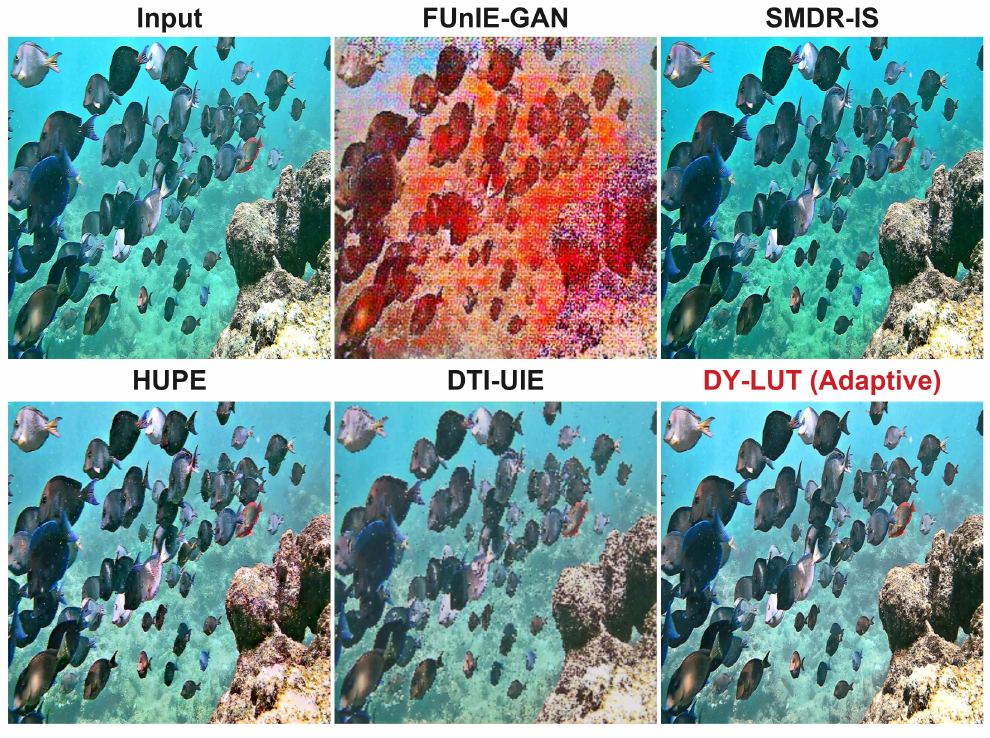}
\caption{Adaptive-inference comparison on native-size 4K UIQAD inputs.}
\label{fig:supp_4k}
\end{figure}

Figures~\ref{fig:supp_1080} and~\ref{fig:supp_4k} show results for native-size 1080P and 4K UIQAD inputs~\cite{chu2023sisc} using adaptive internal-resolution inference. DY-LUT avoids the oversaturation of FUnIE-GAN and the structural softening of DTI-UIE while retaining natural object colors. Bounding the internal workload keeps latency nearly constant across the two resolutions; full-resolution inference remains available when maximum detail is preferred. These images complement the measurements in Tab.~4 of the main paper.

\section{Downstream Visualizations}

The downstream experiments isolate enhancement as the only changed component. Detector weights, feature extractor, thresholds, and geometric verification remain fixed across methods.

\subsection{RUOD Object Detection}

\begin{figure}[!t]
\centering
\includegraphics[width=\columnwidth]{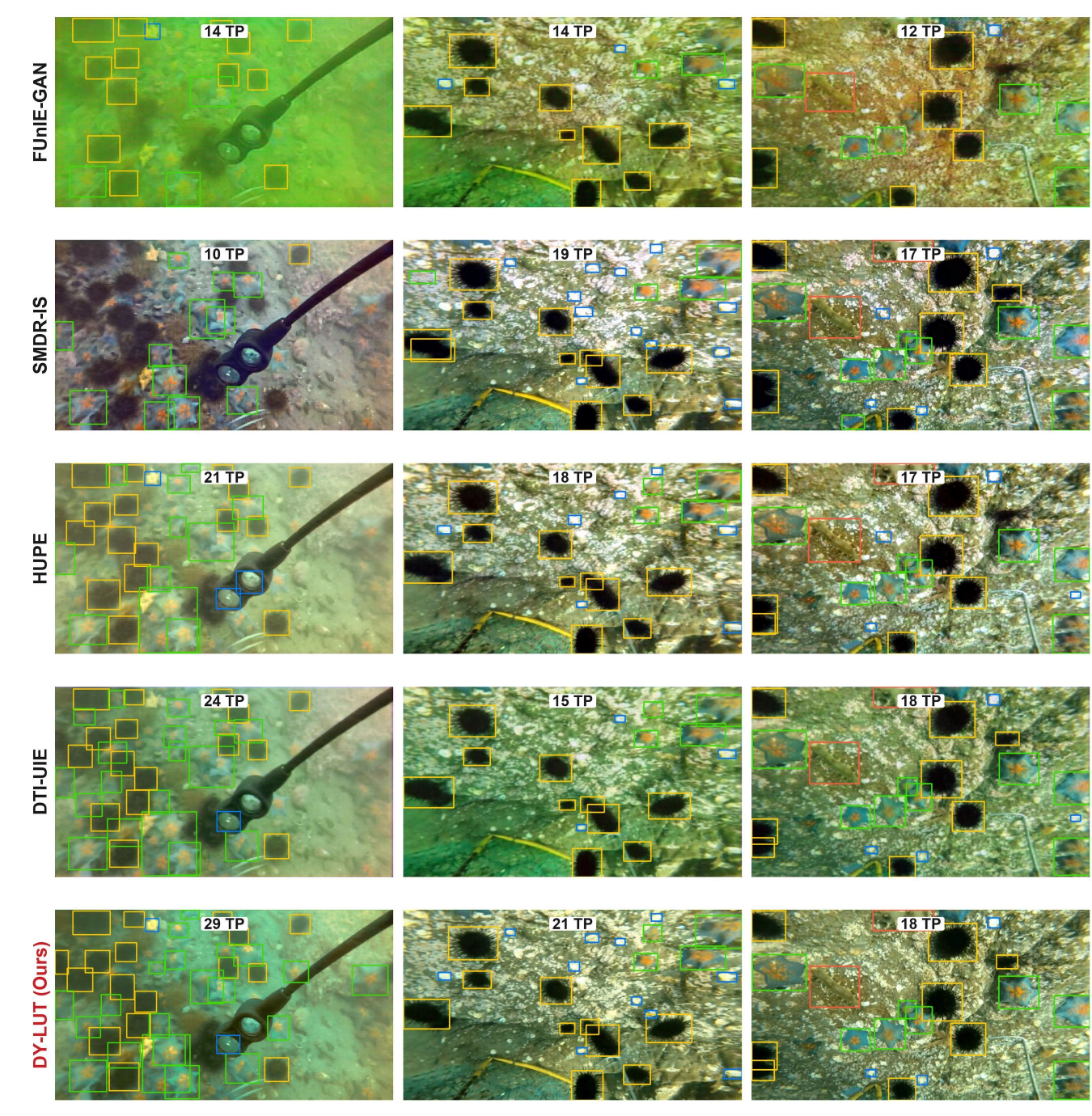}
\caption{YOLO11n detections on RUOD.}
\label{fig:supp_yolo}
\end{figure}

Every enhanced RUOD image is processed by the same fine-tuned YOLO11n detector~\cite{jocher2024yolo11,fu2023rethinking}. Figure~\ref{fig:supp_yolo} shows three representative scenes; the label in each panel reports the number of true-positive detections. DY-LUT improves object visibility, preserves the \mbox{fixed detector}, and produces the largest true-positive count in all three examples.

\subsection{FLSea Feature Matching}

\begin{figure}[!t]
\centering
\includegraphics[width=\columnwidth]{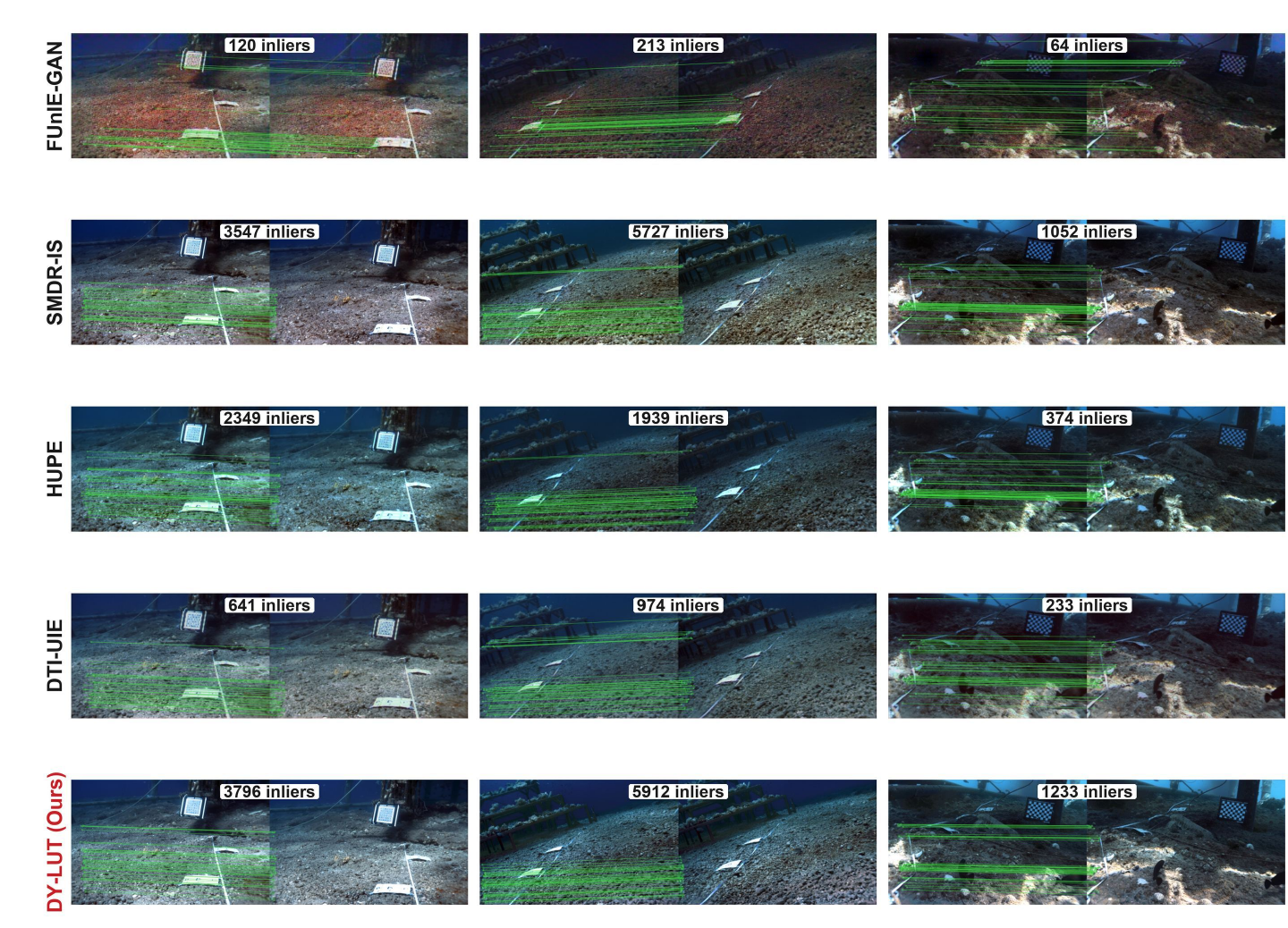}
\caption{Verified SIFT matches on FLSea.}
\label{fig:supp_flsea}
\end{figure}

For FLSea~\cite{randall2023flsea}, every enhanced pair is processed using the same OpenCV SIFT implementation and geometric-verification procedure~\cite{lowe2004distinctive,Bradski2000OpenCV}. Figure~\ref{fig:supp_flsea} displays verified correspondences for three pairs. DY-LUT exposes repeatable structures in low-contrast regions and gives the largest inlier count in each displayed case.

The qualitative task results agree with the averages reported in Tab.~5 of the main paper. In particular, more visible local structure does not automatically imply geometrically reliable matches; the fixed verification step rejects inconsistent correspondences. DY-LUT improves the available feature pool while retaining enough geometric consistency for later perception modules.

\section{Practical Scope}

The figures use one fixed DY-LUT checkpoint for each stated protocol, and no dataset-specific post-processing is applied. Depth is normalized before entering the encoder, so sensor, stereo, or monocular sources can share the same interface. As shown by the main-paper robustness study, retraining can absorb systematic source differences without changing the LUT mechanism.

Across the four enhancement datasets, the comparisons should be read at two complementary scales. Full frames expose global cast removal, luminance balance, and background visibility, whereas aligned crops reveal whether edges, small objects, and fine textures survive the correction. UIEB-90 and LSUI additionally provide references for judging color fidelity; U45 and UIEB-C60 instead test visual plausibility under unpaired and more varied conditions. This consistent behavior supports the use of depth as a spatial cue rather than a dataset-specific appearance prior.

\FloatBarrier

\clearpage
\bibliographystyle{plainnat}
\bibliography{refs}

\end{document}